\documentclass[prd,preprintnumbers,amsmath,amssymb]{revtex4}
\usepackage{graphics,color,subfigure}
\usepackage{epsfig}
\usepackage{dcolumn}
\usepackage{bm}
\usepackage{longtable}

\allowdisplaybreaks
\def\gamf     {\gamma_5}
\def\la{\langle}
\def\ra{\rangle}

\def\chip{\chi_+}

\def\psib{\bar{\psi}}
\def\Psib{\bar{\Psi}}

\def\Bb{\bar{B}}

\def\B{B}

\begin{document}
\title{Relations for low-energy coupling constants in baryon chiral perturbation theory derived from the chiral quark model}

\author{Jun Jiang$^1$}
\author{Shao-Zhou Jiang$^2$}\email{jsz@gxu.edu.cn}
\author{Shi-Yuan Li$^1$}
\author{Yan-Rui Liu$^1$}\email{yrliu@sdu.edu.cn}
\author{Zong-Guo Si$^1$}
\author{Hong-Qian Wang$^1$}

\affiliation{$^1$School of Physics, Shandong University, Jinan, Shandong 250100, China\\
$^2$Key Laboratory for Relativistic Astrophysics, Department of Physics, Guangxi University, Nanning, Guangxi 530004, China}

\date{\today}

\begin{abstract}
The quark model symmetry can be adopted to establish relations between the low-energy constants (LECs) in the baryon chiral perturbation theory ($\chi PT$) if one assumes that a baryon-baryon-meson coupling is described equivalently by a quark-quark-meson coupling at the quark level. Through the correspondence between the $SU(2)$ description and the $SU(3)$ description for the same coupling vertex at the quark level, we find some relations between the LECs in $SU(2)_{\chi PT}$ and $SU(3)_{\chi PT}$ up to the third chiral order. The $SU(3)_{\chi PT}$ LEC relations at the same order are also obtained. The numerical analysis roughly supports these relations. In the situation that the available experimental data are not enough, one may employ such constraints to reduce the number of LECs.
\end{abstract}


\maketitle

\section{Introduction}\label{sec1}

Chiral perturbation theory ($\chi$PT), an effective theory of quantum chromodynamics (QCD) based on the chiral symmetry and its spontaneous breaking, is designed to describe low-energy pion interactions \cite{Weinberg:1978kz,Gasser:1984gg,Gasser:1983yg,Gasser:1987rb}. Both the Lagrangian and amplitude are organized order by order in the chiral expansion (expansion parameter $p$: hadron momentum or pion mass) in this theory. Usually, a consistent power counting scheme is required to decide the needed Lagrangian terms for a calculated amplitude. Since the structures of the interaction are just constrained by symmetries, the number of allowed terms are infinite and one has to introduce an unknown low-energy constant (LEC) in front of each term of the Lagrangian. Such coefficients can be determined phenomenologically before their connections to QCD are clear. In practice, determining the LECs is always a problem in the application of the theory.

Up to now, chiral Lagrangians in meson and baryon sectors have been constructed to high orders. In Refs. \cite{Fearing:1994ga,Herrera-Siklody:1996tqr,Bijnens:1999sh,Bijnens:2001bb,Ebertshauser:2001nj,Cata:2007ns,Haefeli:2007ty,Jiang:2014via,Bijnens:2018lez,Roessl:1999iu,Du:2016xbh,Du:2016ntw}, one can find the chiral Lagrangians in the light meson sector up to the ${\cal O}(p^8)$ order. In the light baryon sector, terms up to the one-loop order have been obtained \cite{Krause:1990xc,Fettes:2000gb,Oller:2006yh,Frink:2006hx,Jiang:2016vax,Jiang:2017yda,Jiang:2018mzd,Holmberg:2018dtv}. In the heavy quark hadron case, both meson \cite{Jenkins:1992hx,Mehen:2005hc,Yao:2015qia,Jiang:2019hgs,Wang:2019mhm,Cheng:1993gc} and baryon \cite{Cheng:1993gc,Lutz:2014jja,Jiang:2014ena,Wang:2018gpl,Wang:2018cre,Heo:2018qnk,Tong:2021raz,Qiu:2020omj} chiral Lagrangians up to the ${\cal O}(p^4)$ order are also explored. The large number of LECs at high orders makes it difficult to achieve accurate high-order calculations in theoretical investigations.

Practically, one may extract the needed LECs from various experiments or lattice simulations \cite{Bali:2022qja,Heo:2019lav,Frezzotti:2021ahg,Aoki:2021kgd,Liu:2006xja,Guo:2012vv}. However, the available data are usually not enough for this purpose when high-order corrections are involved. To reduce the number of unknown parameters, studies from different considerations are essential. One may turn to resonance saturation \cite{Ecker:1988te,Bernard:1993fp,Pich:2008xj,Liu:2010bw,Liu:2011mi,Du:2016tgp}, large $N_c$ \cite{tHooft:1973alw,Witten:1979kh,Dashen:1994qi,Heo:2019cqo,Heo:2022huf}, or quark model \cite{Yan:1992gz,Falk:1992cx}. There are also efforts to determine the LEC values from the fundamental QCD \cite{Wang:1999cp,Yang:2002hea,Chen:2020jiq,Chen:2020epl}. Constraints on the LECs from symmetries, although they cannot give accurate values, are also helpful for us to understand the strong interactions; e.g. to what extent can one understand the hyperon interactions from the nucleon interactions? In the present study, we shall derive some relations between LECs by considering the quark model symmetry.

The flavor symmetry is useful in getting LEC relations. For the pion-nucleon system, one may adopt either $SU(2)$ or $SU(3)$ chiral Lagrangian to describe their interactions. Since the number of terms at the same order is different for these two cases, the relations like $D+F=g_A$ can be derived. Here, $g_A$ is the coefficient of the leading order $\pi NN$ coupling term in the $SU(2)$ baryon chiral perturbation theory and $D$ and $F$ are LECs in the three-flavor version. Since baryons are made of three quarks, more relations are possible if one adopts the chiral quark model ($\chi$QM) \cite{Manohar:1983md,Zhang:1997ny,Dai:2003dz} and the quark wave functions (w.f.) of the baryons. Noticing that the quarks in the $SU(2)$ and $SU(3)$ chiral quark models are all in the fundamental representation and the coupling constants may be equal (called $g_A^q$ in our leading Lagrangians), the hadron-level coupling constants can be related. In the mentioned example, we have $D=\frac35g_A$ and $F=\frac25g_A$. The chiral quark model is just a ``bridge'' for our purpose and the relevant coupling constant $g_A^q$ does not appear in the final expression. The procedure to get such relations is:
\begin{center}
\includegraphics{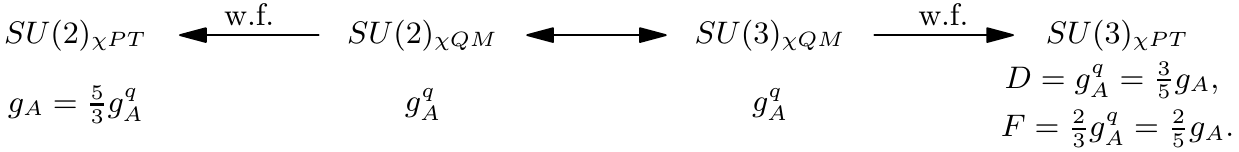}
\end{center}
What we want to do is to find more LEC relations by extending the idea in this example to high-order terms.

This paper is organized as follows. In Sec. \ref{sec2}, we collect all the baryon Lagrangians up to the third chiral order. We also construct relevant Lagrangians at the quark level. In Sec. \ref{sec3}, the nonrelativistic reduction for the coupling terms and the baryon wave functions in the quark model are presented. In Sec. \ref{sec4} and Sec. \ref{sec5}, the relations of coupling constants between $\chi$PT and $\chi$QM in the two-flavor and three-flavor cases are given, respectively. Then we present the coupling constant relations between $SU(2)_{\chi QM}$ and $SU(3)_{\chi QM}$ in Sec. \ref{sec6}. The final LEC relations between $SU(2)_{\chi PT}$ and $SU(3)_{\chi PT}$ are also extracted in Sec. \ref{sec6}. Section \ref{sec7} presents some numerical analyses and Sec. \ref{sec8} is for some discussions.

\section{Lagrangians in $\chi$PT and $\chi$QM}\label{sec2}

\begin{table}[!h]\centering
	\caption{The ${\cal O}(p^1)$ chiral-invariant Lagrangians in $SU(2)_{\chi PT/\chi QM}$, $SU(3)_{\chi QM}$, and $SU(3)_{\chi PT}$. We use $i$ to label the interaction terms and $g_A$, $g_A^q$, and $D/F$ to denote the coupling constants in $SU(2)_{\chi PT}$, $SU(2,3)_{\chi QM}$, and $SU(3)_{\chi PT}$, respectively. $\psi$ in $SU(2)_{\chi QM}$ means $\psi_q$ of Eq. \eqref{defpsiq} and $\langle\cdots\rangle$ means trace in flavor space.}\label{p1}
\begin{tabular}{cc|cc|cccc}\hline\hline
\multicolumn{2}{c|}{$SU(2)_{\chi PT/\chi QM}$}&\multicolumn{2}{c|}{$SU(3)_{\chi QM}$}
&\multicolumn{4}{c}{$SU(3)_{\chi PT}$}\\	
$i$&$g_A/g_A^q$&$i$&$g_A^q$  &  $i$ & $D+F$ & $i$ & $D-F$  \\
1&$\frac12\psib u^{\mu}\gamma_\mu\gamma_5 \psi$ &1 &$\frac12\Psib u^{\mu}\gamma_\mu\gamma_5 \Psi$
&1 & $\frac12\la\Bb u^{\mu}\gamma_\mu\gamma_5  B\ra$ & 2 & $\frac12\la\Bb \gamma_\mu\gamma_5  B u^{\mu}\ra$\\
\hline\hline
\end{tabular}
\end{table}

\begin{table}[!h]\centering
	\caption{The ${\cal O}(p^2)$ chiral-invariant Lagrangians in $SU(2)_{\chi PT/\chi QM}$, $SU(3)_{\chi QM}$, and $SU(3)_{\chi PT}$. We use $i$ to label the interaction terms and $\alpha_i$, $\beta_i$, $c_i$, and $d_i$ to denote the coupling constants in $SU(2)_{\chi PT}$, $SU(2)_{\chi QM}$, $SU(3)_{\chi QM}$, and $SU(3)_{\chi PT}$, respectively. $\psi$ in $SU(2)_{\chi QM}$ means $\psi_q$ of Eq. \eqref{defpsiq} and $\langle\cdots\rangle$ means trace in flavor space. The fifth term in $SU(2)_{\chi PT/\chi QM}$ vanishes because $\la v^\mu\ra=\la a^\mu\ra=0$.}\label{p2}
\begin{tabular}{cc|cc|cccc}\hline\hline
\multicolumn{2}{c|}{$SU(2)_{\chi PT/\chi QM}$}&\multicolumn{2}{c|}{$SU(3)_{\chi QM}$}
&\multicolumn{4}{c}{$SU(3)_{\chi PT}$}\\
$i$&$\alpha_{i}/\beta_{i}$&$i$&$c_i$ &		$i$ & $d_i$ & $i$ & $d_i$  \\
1 & $\psib\la u^{\mu}u_{\mu}\ra\psi$ &1 &$\Psib\la u^{\mu}u_{\mu}\ra\Psi$ & 1 & $\la\Bb u^{\mu}u_{\mu}B\ra$ & 3 & $\la\Bb B u^{\mu}u_{\mu}\ra$\\
  &                                  &2 &$\Psib u^{\mu}u_{\mu}\Psi$ &2 & $\la\Bb u^{\mu}B u_{\mu}\ra$ & 4 & $\la\Bb B\ra\la u^{\mu}u_{\mu}\ra$\\\hline
2 & $i \psib u^{\mu}u^{\nu}\sigma_{\mu\nu}\psi$&3 &$i\Psib u^{\mu}u^{\nu}\sigma_{\mu\nu}\Psi$ & 5 & $i \la\Bb u^{\mu}u^{\nu}\sigma_{\mu\nu}B\ra$ & 7 & $i \la\Bb u^{\mu}\ra\la u^{\nu}\sigma_{\mu\nu}B\ra$  \\
		&&& &6 & $i \la\Bb\sigma_{\mu\nu}B u^{\mu}u^{\nu}\ra$ &  &\\\hline
3 & $\psib\la u^{\mu}u^{\nu}\ra D_{\mu\nu}\psi$&4 &$\Psib\la u^{\mu}u^{\nu}\ra D_{\mu\nu}\Psi$  &  8 & $\la\Bb u^{\mu}u^{\nu}D_{\mu\nu}B \ra$ & 10 & $\la\Bb D_{\mu\nu}B  u^{\mu}u^{\nu}\ra$  \\
		&                                            &5 &$\Psib u^{\mu}u^{\nu}D_{\mu\nu}\Psi$  &  9 & $\la\Bb u^{\mu}D_{\mu\nu}B  u^{\nu}\ra$ & 11 & $\la\Bb D_{\mu\nu}B \ra\la u^{\mu}u^{\nu}\ra$\\\hline
4 & $\psib f_{+}^{\mu\nu}\sigma_{\mu\nu}\psi$ &6&$\Psib f_{+}^{\mu\nu}\sigma_{\mu\nu}\Psi$  &  12 & $\la\Bb f_{+}^{\mu\nu}\sigma_{\mu\nu}B\ra$ & 13 &$\la\Bb\sigma_{\mu\nu}B f_{+}^{\mu\nu}\ra$ \\
5 & $\psib\la f_{+}^{\mu\nu}\ra\sigma_{\mu\nu}\psi\to 0$&&  \\\hline
$6$ & $\psib\tilde{\chi}_+\psi$  &7&$\Psib\tilde{\chi}_+\Psi$  &  14 & $\la\Bb\tilde{\chi}_+ B\ra$ & 16 & $\la\Bb B\ra\la\chip\ra$\\
$7$ & $\psib\la\chip\ra\psi$  &8&$\Psib\la\chip\ra\Psi$  & 15 & $\la\Bb B\tilde{\chi}_+\ra$\\
\hline\hline
\end{tabular}
\end{table}

\begin{table}[!h]\centering
\caption{The ${\cal O}(p^3)$ chiral-invariant Lagrangians in $SU(2)_{\chi PT/\chi QM}$, $SU(3)_{\chi QM}$, and $SU(3)_{\chi PT}$. We use $i$ to label the interaction terms and $\alpha_i$, $\beta_i$, $c_i$, and $d_i$ to denote the coupling constants in $SU(2)_{\chi PT}$, $SU(2)_{\chi QM}$, $SU(3)_{\chi QM}$, and $SU(3)_{\chi PT}$, respectively. $\psi$ in $SU(2)_{\chi QM}$ means $\psi_q$ of Eq. \eqref{defpsiq} and $\langle\cdots\rangle$ means trace in flavor space. The symbol $[\cdots]_+$ stands for $[\cdots]+\mathrm{H.c.}$. The tenth and 13th terms in $SU(2)_{\chi PT/\chi QM}$ vanish because $\la v^\mu\ra=\la a^\mu\ra=0$.}\label{p3}
\begin{tabular}{cc|cc|cccc}\hline\hline
\multicolumn{2}{c|}{$SU(2)_{\chi PT/\chi QM}$}&\multicolumn{2}{c|}{$SU(3)_{\chi QM}$}
&\multicolumn{4}{c}{$SU(3)_{\chi PT}$}\\
$i$&$\alpha_{i}/\beta_{i}$&$i$&$c_i$ & $i$ &$d_i$& $i$ &$d_i$ \\
	
	1 & $\psib\la u^{\mu}u_{\mu}\ra u^{\nu}\gamma_{\nu} \gamf              \psi$ & 1&$\Psib\la u^{\mu}u_{\mu}\ra u^{\nu}\gamma_{\nu} \gamf              \Psi$ & 1 & $[\la\Bb u^{\mu}u_{\mu}u^{\nu}\gamma_{\nu} \gamf B   \ra]_+$&7 &$\la\Bb\gamma_{\nu} \gamf      B u^{\mu}u^{\nu}u_{\mu}\ra$ \\
	2 & $\psib\la u^{\mu}u^{\nu}\ra u_{\mu}\gamma_{\nu} \gamf              \psi$ & 2&$\Psib\la u^{\mu}u^{\nu}\ra u_{\mu}\gamma_{\nu} \gamf              \Psi$ &2 &$\la\Bb u^{\mu}u^{\nu}u_{\mu}\gamma_{\nu} \gamf B   \ra$  &8 &$[\la\Bb\gamma_{\nu} \gamf      B u^{\nu}u^{\mu}u_{\mu}\ra]_+$ \\
	&&3&$\Psib\la u^{\mu}u_{\mu}u^{\nu}\ra\gamma_{\nu} \gamf              \Psi$&3 & $\la\Bb u^{\mu}u_{\mu}\gamma_{\nu} \gamf B    u^{\nu}\ra$ &9 &$\la\Bb\gamma_{\nu} \gamf B    u^{\nu}\ra\la u^{\mu}u_{\mu}\ra$ \\
	&&4&$[\Psib u^{\mu}u_{\mu}u^{\nu}\gamma_{\nu} \gamf              \Psi]_+$&4 & $[\la\Bb u^{\nu}u^{\mu}\gamma_{\nu} \gamf      B u_{\mu}\ra]_+$ &10& $\la\Bb\gamma_{\nu} \gamf      B\ra\la u^{\nu}u^{\mu}u_{\mu}\ra$\\
	&&&&5 & $\la\Bb u^{\nu}\gamma_{\nu} \gamf B u^{\mu}u_{\mu}\ra$&11& $[\la\Bb u^{\mu}u_{\mu}\ra\la u^{\nu}\gamma_{\nu} \gamf B\ra]_+$\\
	&&&&6 & $[\la\Bb u_{\mu}\gamma_{\nu} \gamf B u^{\mu}u^{\nu}\ra]_+$\\\hline
	3 & $\varepsilon_{\mu\nu\lambda\rho}\psib\la u^{\mu}u^{\nu}u^{\lambda}\ra D^{\rho}\psi$
	& 5&$\epsilon_{\mu\nu\lambda\rho}\Psib\la u^{\mu}u^{\nu}u^{\lambda}\ra D^{\rho}\Psi$ & 12 & $\varepsilon_{\mu\nu\lambda\rho}\la\Bb u^{\mu}u^{\nu}u^{\lambda}D^{\rho}B \ra$ & 15& $\varepsilon_{\mu\nu\lambda\rho}\la\Bb D^{\rho}B  u^{\mu}u^{\nu}u^{\lambda}\ra$ \\
	&&6&$\epsilon_{\mu\nu\lambda\rho}\Psib u^{\mu}u^{\nu}u^{\lambda}D^{\rho}\Psi$&13& $\varepsilon_{\mu\nu\lambda\rho}\la\Bb u^{\mu}u^{\nu}D^{\rho}B  u^{\lambda}\ra$& 16& $\varepsilon_{\mu\nu\lambda\rho}\la\Bb D^{\rho}B \ra\la u^{\mu}u^{\nu}u^{\lambda}\ra$\\
	&&&&14& $\varepsilon_{\mu\nu\lambda\rho}\la\Bb u^{\lambda}D^\rho B  u^{\mu}u^{\nu}\ra$\\\hline
	4 & $\psib\la u^{\mu}u^{\nu}\ra u^{\lambda}\gamma_{\mu} \gamf      D_{\nu\lambda}\psi$ & 7 & $\Psib\la u^{\mu}u^{\nu}\ra u^{\lambda}\gamma_{\mu} \gamf      D_{\nu\lambda}\Psi$ & 17& $[\la\Bb u^{\mu}u^{\nu}u^{\lambda}\gamma_{\mu} \gamf      D_{\nu\lambda}B \ra]_+$& 23 & $[\la\Bb\gamma_{\mu} \gamf      D_{\nu\lambda}B  u^{\mu}u^{\nu}u^{\lambda}\ra]_+$\\
	5 & $\psib\la u^{\mu}u^{\nu}\ra u^{\lambda}\gamma_{\lambda} \gamf       D_{\mu\nu}\psi$ & 8 & $\Psib\la u^{\nu}u^{\lambda}\ra u^{\mu}\gamma_{\mu} \gamf D_{\nu\lambda}\Psi$ & 18 & $\la\Bb u^{\nu}u^{\mu}u^{\lambda}\gamma_{\mu} \gamf D_{\nu\lambda}B \ra$& 24 & $\la\Bb\gamma_{\mu} \gamf      D_{\nu\lambda}B  u^{\nu}u^{\mu}u^{\lambda}\ra$\\
	&&9&$\Psib\la u^{\mu}u^{\nu}u^{\lambda}\ra\gamma_{\mu} \gamf      D_{\nu\lambda}\Psi$&19 & $\la\Bb u^{\nu}u^{\lambda}\gamma_{\mu} \gamf       D_{\nu\lambda}B  u^{\mu}\ra$& 25 & $\la\Bb\gamma_{\mu} \gamf      D_{\nu\lambda}B \ra\la u^{\mu}u^{\nu}u^{\lambda}\ra$\\
	&&10&$[\Psib u^{\mu}u^{\nu}u^{\lambda}\gamma_{\mu} \gamf      D_{\nu\lambda}\Psi]_+$&20& $[\la\Bb u^{\mu}u^{\nu}\gamma_{\mu} \gamf      D_{\nu\lambda}B  u^{\lambda}\ra]_+$ & 26 & $\la\Bb\gamma_{\mu} \gamf      D_{\nu\lambda}B  u^{\mu}\ra\la u^{\nu}u^{\lambda}\ra$ \\
	&&&&21& $[\la\Bb u^{\lambda}\gamma_{\mu} \gamf D_{\nu\lambda}B  u^{\mu}u^{\nu}\ra]_+$ & 27 & $[\la\Bb u^{\nu}u^{\lambda}\ra\la u^{\mu}\gamma_{\mu} \gamf D_{\nu\lambda}B \ra]_+$\\
	&&&&22& $\la\Bb u^{\mu}\gamma_{\mu} \gamf      D_{\nu\lambda}B  u^{\nu}u^{\lambda}\ra$\\\hline
	
	6 & $[\psib u_{\mu}h^{\mu\nu}D_{\nu}\psi]_+$ & 11&$[\Psib u_{\mu}h^{\mu\nu}D_{\nu}\Psi]_+$ & 28 & $[\la\Bb u_{\mu}h^{\mu\nu}D_{\nu}B \ra]_+$&30& $[\la\Bb h^{\mu\nu}\ra\la u_{\mu}D_{\nu}B \ra]_+$ \\
	&&&&29& $[\la\Bb D_{\nu}B  u_{\mu}h^{\mu\nu}\ra]_+$\\\hline
	7 & $[\psib u^{\mu}h^{\nu\lambda}D_{\mu\nu\lambda}\psi]_+$ & 12&$[\Psib u^{\mu}h^{\nu\lambda}D_{\mu\nu\lambda}\Psi]_+$ &31 & $[\la\Bb u^{\mu}h^{\nu\lambda}D_{\mu\nu\lambda}B \ra]_+$&33 & $[\la\Bb h^{\mu\nu}\ra\la u^{\lambda}D_{\mu\nu\lambda}B \ra]_+$\\
	&&&&32 & $[\la\Bb D_{\mu\nu\lambda}B  u^{\mu}h^{\nu\lambda}\ra]_+$\\\hline
	
	8 & $i \psib\la u^{\mu}h^{\nu\lambda}\ra\sigma_{\mu\nu}D_{\lambda}\psi$ & 13&$i\Psib\la u^{\mu}h^{\nu\lambda}\ra\sigma_{\mu\nu}D_{\lambda}\Psi$ & 34 & $[i \la\Bb u^{\mu}h^{\nu\lambda}\sigma_{\mu\nu}D_{\lambda}B \ra]_+$ & 36& $[i \la\Bb\sigma_{\mu\nu}D_{\lambda}B  u^{\mu}h^{\nu\lambda}\ra]_+$\\
	&&14&$[i\Psib u^{\mu}h^{\nu\lambda}\sigma_{\mu\nu}D_{\lambda}\Psi]_+$& 35& $i \la\Bb u^{\mu}\sigma_{\mu\nu}D_{\lambda}B  h^{\nu\lambda}\ra$& 37& $i \la\Bb\sigma_{\mu\nu}D_{\lambda}B \ra\la u^{\mu}h^{\nu\lambda}\ra$\\\hline
	9 & $[i \psib f_{+}^{\mu\nu}u_{\mu}\gamma_{\nu} \gamf              \psi]_+$ & 15&$[i\Psib f_{+}^{\mu\nu}u_{\mu}\gamma_{\nu} \gamf              \Psi]_+$& 38 & $[i \la\Bb f_{+}^{\mu\nu}u_{\mu}\gamma_{\nu} \gamf B   \ra]_+$ & 40 & $[i \la\Bb u_{\mu}\ra\la f_+^{\mu\nu}\gamma_{\nu} \gamf B   \ra]_+$ \\
	&&&& 39& $[i \la\Bb\gamma_{\nu} \gamf Bf_+^{\mu\nu}u_{\mu}\ra]_+$\\
	\hline
	10 & $i \varepsilon_{\mu\nu\lambda\rho}\psib\la f_{+}^{\mu\nu}\ra u^{\lambda}D^{\rho}\psi\to 0$ &16 &$i\epsilon_{\mu\nu\lambda\rho}\Psib\la f_{+}^{\mu\nu}u^{\lambda}\ra D^{\rho}\Psi$& 41 & $[i \varepsilon_{\mu\nu\lambda\rho}\la\Bb f_+^{\mu\nu}u^{\lambda}D^{\rho}B \ra]_+$& 44  & $[i \varepsilon_{\mu\nu\lambda\rho}\la\Bb D^{\rho}B  f_+^{\mu\nu}u^{\lambda}\ra]_ +$\\
	11 & $i \varepsilon_{\mu\nu\lambda\rho}\psib\la f_{+}^{\mu\nu}u^{\lambda}\ra D^{\rho}\psi$ &17&$[i\epsilon_{\mu\nu\lambda\rho}\Psib f_+^{\mu\nu}u^{\lambda}D_{\rho}\Psi]_+$&42 & $i \varepsilon_{\mu\nu\lambda\rho}\la\Bb f_+^{\mu\nu}D^{\rho}B  u^{\lambda}\ra$& 45 & $i \varepsilon_{\mu\nu\lambda\rho}\la\Bb D^{\rho}B \ra\la f_+^{\mu\nu}u^{\lambda}\ra$ \\
	&&&&43& $i \varepsilon_{\mu\nu\lambda\rho}\la\Bb u^{\lambda}D^{\rho}B  f_+^{\mu\nu}\ra$\\
	\hline
	
	12 & $i \psib\nabla_{\mu}f_{+}^{\mu\nu}D_{\nu}\psi$& 18&$i\Psib\nabla_{\mu}f_+^{\mu\nu}D_{\nu}\Psi$ & 46& $i \la\Bb\nabla_{\mu}f_+^{\mu\nu}D_{\nu}B \ra$ & 47& $i \la\Bb D_{\nu}B \nabla_{\mu}f_+^{\mu\nu}\ra$\\
	13 & $i \psib\la\nabla_{\mu}f_+^{\mu\nu}\ra D_{\nu}\psi\to 0$&& \\\hline
	14 & $[i \psib f_{+}^{\mu\nu}u^{\lambda}\gamma_{\mu} \gamf      D_{\nu\lambda}\psi]_+$ & 19 & $[i\Psib f_{+}^{\mu\nu}u^{\lambda}\gamma_{\mu} \gamf      D_{\nu\lambda}\Psi]_+$ & 48 & $[i \la\Bb f_{+}^{\mu\nu}u^{\lambda}\gamma_{\mu} \gamf      D_{\nu\lambda}B \ra]_+$ &50 & $[i \la\Bb u^{\lambda}\ra\la f_{+}^{\mu\nu}\gamma_{\mu} \gamf D_{\nu\lambda}B \ra]_+$\\
	&&&& 49   & $[i \la\Bb\gamma_{\mu} \gamf      D_{\nu\lambda}B  f_+^{\mu\nu}u^{\lambda}\ra ]_+$\\\hline
	15 & $[\psib u_{\mu}f_{-}^{\mu\nu}D_{\nu}\psi]_+$ &  20 & $[\Psib u_{\mu}f_-^{\mu\nu}D_{\nu}\Psi]_+$
	& 51& $[\la\Bb u_{\mu}f_-^{\mu\nu}D_{\nu}B \ra ]_+$ & 53& $[\la\Bb f_{-}^{\mu\nu}\ra\la u_{\mu}D_{\nu}B \ra]_+$\\
	&&&& 52& $[\la\Bb D_{\nu}B  u_{\mu}f_-^{\mu\nu}\ra]_+$\\\hline
	16 & $i \psib\la u^{\mu}f_{-}^{\nu\lambda}\ra\sigma_{\mu\nu}D_{\lambda}\psi$ & 21& $i\Psib\la u^{\mu}f_{-}^{\nu\lambda}\ra\sigma_{\mu\nu}D_{\lambda}\Psi$ & 54 & $[i \la\Bb u^{\mu}f_{-}^{\nu\lambda}\sigma_{\mu\nu}D_{\lambda}B \ra]_+$& 59 & $i \la\Bb u^{\mu}\sigma_{\nu\lambda}D_{\mu}B  f_-^{\nu\lambda}\ra$\\
	17 & $i \psib\la u^{\mu}f_{-}^{\nu\lambda}\ra\sigma_{\nu\lambda}D_{\mu}\psi$ & 22 & $i\Psib\la u^{\mu}f_{-}^{\nu\lambda}\ra\sigma_{\nu\lambda}D_{\mu}\Psi$ & 55 & $[i \la\Bb u^{\mu}f_{-}^{\nu\lambda}\sigma_{\nu\lambda}D_{\mu}B \ra]_+$ & 60 & $[i \la\Bb\sigma_{\mu\nu}D_{\lambda}B  u^{\mu}f_-^{\nu\lambda}\ra ]_+$\\
	&&23&$[i\Psib u^{\mu}f_{-}^{\nu\lambda}\sigma_{\mu\nu}D_{\lambda}\Psi]_+$& 56 & $i \la\Bb f_{-}^{\nu\lambda}\sigma_{\nu\lambda}D_{\mu}B  u^{\mu}\ra$ & 61 & $[i \la\Bb\sigma_{\nu\lambda}D_{\mu}B  u^{\mu}f_-^{\nu\lambda}\ra ]_+$\\
	&&24&$[i\Psib u^{\mu}f_{-}^{\nu\lambda}\sigma_{\nu\lambda}D_{\mu}\Psi]_+$& 57 & $i \la\Bb f_{-}^{\nu\lambda}\sigma_{\mu\nu}D_{\lambda}B  u^{\mu}\ra$ & 62 & $i \la\Bb\sigma_{\mu\nu}D_{\lambda}B \ra\la u^{\mu}f_-^{\nu\lambda}\ra$\\
	&&&&58&$i \la\Bb u^{\mu}\sigma_{\mu\nu}D_{\lambda}B  f_-^{\nu\lambda}\ra$ &63&$i \la\Bb\sigma_{\nu\lambda}D_{\mu}B \ra\la u^{\mu}f_-^{\nu\lambda}\ra$
	\\\hline
	18 & $\psib\nabla_{\mu}f_{-}^{\mu\nu}\gamma_{\nu} \gamf              \psi$ & 25 & $\Psib\nabla_{\mu}f_-^{\mu\nu}\gamma_{\nu} \gamf              \Psi$
	& 64& $\la\Bb\nabla_{\mu}f_-^{\mu\nu}\gamma_{\nu} \gamf B   \ra$ &65 & $\la\Bb\gamma_{\mu} \gamf      B\nabla_{\nu}f_-^{\mu\nu}\ra$ \\\hline
	$19$ & $\psib\la u^{\mu}\tilde{\chi}_+\ra\gamma_{\mu} \gamf      \psi$ &26 & $\Psib\la u^{\mu}\tilde{\chi}_+\ra\gamma_{\mu} \gamf      \Psi$ & 66 & $[\la\Bb u^{\mu}\tilde{\chi}_+\gamma_{\mu} \gamf      B\ra]_+$ & 70 & $\la\Bb\gamma_{\mu} \gamf      B\ra\la u^{\mu}\tilde{\chi}_+\ra$\\
	$20$ & $\psib\la\chip\ra u^{\mu}\gamma_{\mu} \gamf      \psi$ &27 & $\Psib\la\chip\ra u^{\mu}\gamma_{\mu} \gamf      \Psi$& 67 & $\la\Bb\tilde{\chi}_+\gamma_{\mu} \gamf      B u^{\mu}\ra$ &  71 &$\la\Bb u^{\mu}\gamma_{\mu} \gamf      B \ra\la\chip\ra$ \\
	&&28&$[\Psib u^{\mu}\tilde{\chi}_+\gamma_{\mu} \gamf      \Psi]_+$&68 & $\la\Bb u^{\mu}\gamma_{\mu} \gamf      B\tilde{\chi}_+\ra$ & 72 & $\la\Bb\gamma_{\mu} \gamf      B u^{\mu}\ra\la\chip\ra$\\
	&&&&69 & $[\la\Bb\gamma_{\mu} \gamf      B u^{\mu}\tilde{\chi}_+\ra]_+$& &\\\hline
	$21$ & $i \psib\tilde{\chi}_-^\mu\gamma_\mu \gamf \psi$    &29 & $i\Psib\tilde{\chi}_-^\mu\gamma_\mu \gamf \Psi$ & 73 & $i \la\Bb\tilde{\chi}_-^\mu\gamma_\mu \gamf  B\ra$ &75 & $i \la\Bb\gamma_{\mu} \gamf  B\ra\la\chi_-^\mu\ra$ \\
	$22$ & $i \psib\la\chi_{-}^{\mu}\ra\gamma_{\mu} \gamf      \psi$ &30 &$i\Psib\la\chi_{-}^{\mu}\ra\gamma_{\mu} \gamf      \Psi$ & 74 & $i \la\Bb\gamma_{\mu} \gamf      B\tilde{\chi}_-^\mu\ra$\\\hline
	23 & $[i \psib u^{\mu}\tilde{\chi}_- D_\mu\psi]_+$ &31&$[i\Psib u^{\mu}\tilde{\chi}_- D_\mu\Psi]_+$ &  76 & $[i \la\Bb u^{\mu}\tilde{\chi}_- D_{\mu}B \ra]_+$& 78 & $[i \la\Bb\tilde{\chi}_-\ra\la u^{\mu}D_{\mu}B \ra]_+$\\
	&&&& 77 & $[i \la\Bb D_{\mu}B  u^{\mu}\tilde{\chi}_-\ra]_+$\\
	\hline\hline
\end{tabular}
\end{table}

To get the LEC relations, we need the complete chiral Lagrangians and classify the terms based on their structures. According to the power counting rules, the transformation properties (charge conjugation, parity, and Hermiticity) of various building blocks, and the reduction procedure, one can obtain the minimal chiral-invariant Lagrangians. The explicit forms of the interaction terms up to ${\cal O}(p^4)$ have been given in Refs. \cite{Fettes:2000gb,Oller:2006yh,Frink:2006hx,Jiang:2016vax}. Here, we restrict our discussions up to the third chiral order and we collect the adopted terms in Tables \ref{p1}, \ref{p2}, and \ref{p3}. They are classified into 1, 5, and 16 groups separated by hlines for ${\cal O}(p^1)$, ${\cal O}(p^2)$, and ${\cal O}(p^3)$ Lagrangians, respectively. The first (third) columns list terms in the two-flavor (three-flavor) baryon $\chi$PT. The involved baryon fields are
\begin{eqnarray}
\psi=\left(\begin{array}{c}p\\n\end{array}\right),
\qquad B=\left(\begin{array}{ccc}\dfrac{\Sigma^0}{\sqrt{2}}+\dfrac{\Lambda}{\sqrt{6}}&\Sigma^+&p\\\Sigma^-&-\dfrac{\Sigma^0}{\sqrt{2}}+\dfrac{\Lambda}{\sqrt{6}}&n\\\Xi^-&\Xi^0&-\dfrac{2\Lambda}{\sqrt{6}}\end{array}\right).
\end{eqnarray}
Other definitions are \cite{Fettes:2000gb,Oller:2006yh,Frink:2006hx,Jiang:2016vax}
\begin{eqnarray}
u^\mu&=&i\{u^\dag(\partial^\mu-ir^\mu)u-u(\partial^\mu-il^\mu)u^\dag\},\nonumber\\
u&=&\exp(\frac{i\phi}{2f_\pi}), \quad \phi=\pi^i\lambda^i \text{ or }\pi^i\tau^i,\nonumber\\
r^\mu&=&v^\mu+a^\mu,\quad l^\mu=v^\mu-a^\mu,\nonumber\\
\chi_{\pm}&=&u^\dag\chi u^\dag\pm u\chi^\dag u, \quad \chi=2B_0(s+ip)\nonumber\\
h^{\mu\nu}&=&\nabla^\mu u^\nu+\nabla^\nu u^\mu,\nonumber\\
f_+^{\mu\nu}&=&u F_L^{\mu\nu} u^\dag+u^\dag F_R^{\mu\nu} u,\nonumber\\
f_-^{\mu\nu}&=&u F_L^{\mu\nu} u^\dag-u^\dag F_R^{\mu\nu} u=-\nabla^\mu u^\nu+\nabla^\nu u^\mu,\nonumber\\
F_L^{\mu\nu}&=&\partial^{\mu}l^{\nu}-\partial^{\nu}l^{\mu}-i[l^{\mu},l^{\nu}],\nonumber\\
F_R^{\mu\nu}&=&\partial^{\mu}r^{\nu}-\partial^{\nu}r^{\mu}-i[r^{\mu},r^{\nu}],\nonumber\\
\Gamma^\mu&=&\frac12\{u^\dag(\partial^\mu-ir^\mu)u+u(\partial^\mu-il^\mu)u^\dag\},\nonumber\\
\chi^\mu_\pm&=&u^\dag\tilde{\nabla}^\mu\chi u^\dag \pm u\tilde{\nabla}^\mu\chi^\dag u =\nabla^\mu\chi_\pm-\frac{i}{2}\{\chi_{\mp},u^\mu\},\nonumber\\
\nabla^\mu O&=&\partial^\mu O+[\Gamma^\mu,O],\nonumber\\
\tilde{\nabla}^\mu\chi&=&\partial^\mu\chi-ir^\mu\chi+i\chi l^\mu,\nonumber\\
D^\mu B&=&\partial^\mu B+[\Gamma^\mu,B],\nonumber\\
D^\mu \psi&=&\partial^\mu\psi+\Gamma^\mu\psi,\nonumber\\
D_{\nu\lambda\rho\cdots}&=&D_\nu D_\lambda D_\rho\cdots+\text{full permutation of }D\text{'s}.
\end{eqnarray}

Here, $\lambda^i$ $(i=1,2,\cdots,8)$ are the Gell-Mann matrices, $\tau^i$ $(i=1,2,3)$ are the Pauli matrices, and $B_0$ is a constant related to the quark condensate. For the scalar ($s$), pseudoscalar ($p$), vector ($v_\mu$), and axial-vector ($a_\mu$) external sources, we take $\langle v_\mu\rangle=\langle a_\mu\rangle=0$. The resulting terms (the fifth term in table \ref{p2}, the tenth and 13th terms in Table \ref{p3}) related to $\langle f_+^{\mu\nu}\rangle$ in $SU(2)_{\chi PT}$ would vanish and could actually be removed. For convenience \cite{Fettes:2000gb}, we also separate the matrix $\chi_+$ into two parts: the trace part $\langle\chi_+\rangle$ and the traceless part $\tilde{\chi}_+=\chi_+-\frac{1}{N_f}\langle\chi_+\rangle$ with $N_f$ being the number of flavor. The matrices $\chi_-$ and $\chi_\pm^\mu$ are treated similarly.

In the Lagrangians, we use $g_A$ and $D/F$ to denote the leading order coupling coefficients in the $SU(2)$ $\chi$PT and $SU(3)$ $\chi$PT, respectively. For high-order terms, here we simply use the symbols $\alpha_i$ and $d_i$ to denote the LECs in $SU(2)$ and $SU(3)$ $\chi$PTs, respectively. The terms in the tables are classified into different groups so that the items in the same group contribute to the same vertex structure.

\begin{figure}[htbp]
\centering
\includegraphics{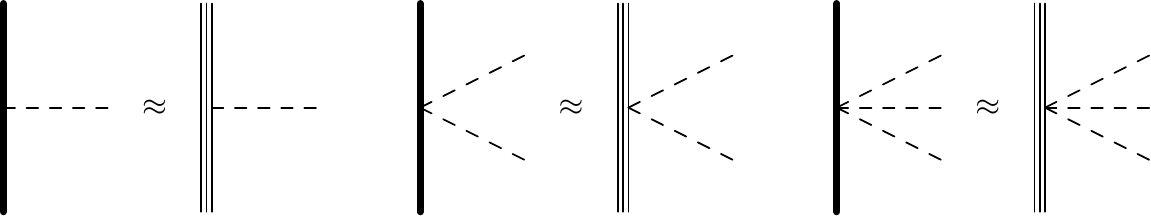}
\caption{Schematic figure for the approximation we adopt. A baryon-baryon-meson coupling in $\chi$PT is approximately described by a quark-quark-meson coupling at the quark level. Two quarks in the baryons are treated as spectators.}\label{chiQM}
\end{figure}

In principle, the matrix of a coupling vertex can be calculated at both hadron and quark levels. When adopting the $\chi$QM to find LEC relations, we use the approximations illustrated in Fig. \ref{chiQM}. We assume that the coupling of a baryon with the pions is due to the coupling of one quark with the external pions, i.e. two quarks are just spectators. Therefore, we also need the high-order coupling terms in the $\chi$QM. In a similar way to construct the pion-nucleon Lagrangians, the perturbative Lagrangians in the $\chi$QM can be obtained. We also list them in the Tables \ref{p1}, \ref{p2}, and \ref{p3}. Since both nucleons and the $u$ and $d$ quarks are in the fundamental representation of the flavor group $SU(2)$, the terms in $SU(2)_{\chi QM}$ and $SU(2)_{\chi PT}$ have the same forms, but the matter field is now
\begin{eqnarray}\label{defpsiq}
\psi\to \psi_q=\left(\begin{array}{c}u\\d\end{array}\right).
\end{eqnarray}
Compared with the two-flavor case, the $SU(3)_{\chi QM}$ has more structures at ${\cal O}(p^2)$ and ${\cal O}(p^3)$. They are illustrated in the second columns of Tables \ref{p1}, \ref{p2}, and \ref{p3}, where
\begin{eqnarray}
\Psi=\left(\begin{array}{c}u\\d\\s\end{array}\right).
\end{eqnarray}
In the leading order Lagrangians, there is only one coupling term in both $SU(2)_{\chi QM}$ and $SU(3)_{\chi QM}$. The coupling constants are the same and we use $g_A^q$ to denote it. For high-order terms, we use $\beta_i$ and $c_i$ to denote the coupling constants in $SU(2)_{\chi QM}$ and $SU(3)_{\chi QM}$, respectively. For convenience, we also call them LECs.

\section{Nonrelativistic reduction and baryon wave functions}\label{sec3}

Employing baryon wave functions to find LEC relations means that we need to reduce the covariant formalism of the Lagrangian to the nonrelativistic form. From Tables \ref{p1}, \ref{p2}, and \ref{p3}, there are three Lorentz structures we should deal with. Their reductions read
\begin{eqnarray}\label{1.1}
\bar{\psi}\psi&\to& \psi_H^\dag \psi_H;\notag\\
\bar{\psi}\gamma^\mu\gamma_5\psi\approx 2\bar{\psi}_vS^\mu \psi_v&\to&
\left\{\begin{array}{cc}0,&(\mu=0)\notag\\ \psi_H^\dag\sigma^k \psi_H,&(\mu=k)\end{array}\right.;\notag\\
\bar{\psi}\sigma^{\mu\nu}\psi\approx 2\epsilon^{\mu\nu\alpha\beta}v_\alpha\bar{\psi}_v S_\beta\psi_v&\to&\left\{\begin{array}{cc}\epsilon^{ijk}\psi_H^\dag\sigma^k \psi_H,&(\mu=i,\nu=j)\\0,&(other)\end{array}\right.,
\end{eqnarray}
where $S^\mu=\frac{i}{2}\gamma_5\sigma^{\mu\nu}v_\nu$, $\psi_v=\frac{1+v\!\!\!/}{2}\psi$ ($v^\mu$ is the baryon 4-velocity), and $\psi_H$ is the large component of $\psi$. With the reduced baryon-meson interaction term, it is easy to get the coupling matrix element at the hadron level. Adopting the approximation shown in Fig. \ref{chiQM} and the quark wave functions of the baryons, one can calculate the same matrix element at the quark level. Then the relation between the coupling coefficients in $\chi PT$ and $\chi QM$ follows from matching. Details will be given later.

In the quark model, the octet baryon flavor-spin wave functions are \cite{Close:1979bt}
\begin{eqnarray}\label{PsiFS}
\Psi_{FS}=\frac{1}{\sqrt2}(\phi^{MS}\otimes\chi^{MS}+\phi^{MA}\otimes\chi^{MA})
\end{eqnarray}
where the flavor wave functions $\phi^{MS}$ and $\phi^{MA}$ and the spin wave functions $\chi^{MS}$ and $\chi^{MA}$ are constructed with the $SU(3)$ and $SU(2)$ Clebsch-Gordan coefficients \cite{deSwart:1963pdg,Kaeding:1995vq}, respectively. For our purpose, one may just focus on the spin-up baryon states in the following calculation. Explicitly, these wave functions read
\begin{eqnarray}\label{wf-octet}
p_\uparrow&=&\frac{1}{3\sqrt2} [udu\otimes(2\uparrow\downarrow\uparrow-\downarrow\uparrow\uparrow-\uparrow\uparrow\downarrow)+duu\otimes(2\downarrow\uparrow\uparrow-\uparrow\downarrow\uparrow-\uparrow\uparrow\downarrow) -uud\otimes(\uparrow\downarrow\uparrow+ \downarrow\uparrow\uparrow-2\uparrow\uparrow\downarrow)],\nonumber\\
n_\uparrow&=&\frac{1}{3\sqrt2} [udd\otimes(\uparrow\downarrow\uparrow-2\downarrow\uparrow\uparrow+\uparrow\uparrow\downarrow)+dud\otimes(\downarrow\uparrow\uparrow-2\uparrow\downarrow\uparrow+\uparrow\uparrow\downarrow) +ddu\otimes(\uparrow\downarrow\uparrow+ \downarrow\uparrow\uparrow-2\uparrow\uparrow\downarrow)],\nonumber\\
\Sigma^+_\uparrow&=&-\frac{1}{3\sqrt2} [usu\otimes(2\uparrow\downarrow\uparrow-\downarrow\uparrow\uparrow-\uparrow\uparrow\downarrow)+suu\otimes(2\downarrow\uparrow\uparrow-\uparrow\downarrow\uparrow-\uparrow\uparrow\downarrow) -uus\otimes(\uparrow\downarrow\uparrow+ \downarrow\uparrow\uparrow-2\uparrow\uparrow\downarrow)],\nonumber\\
\Sigma^-_\uparrow&=&\frac{1}{3\sqrt2} [dsd\otimes(2\uparrow\downarrow\uparrow-\downarrow\uparrow\uparrow-\uparrow\uparrow\downarrow)+sdd\otimes(2\downarrow\uparrow\uparrow-\uparrow\downarrow\uparrow-\uparrow\uparrow\downarrow) -dds\otimes(\uparrow\downarrow\uparrow+ \downarrow\uparrow\uparrow-2\uparrow\uparrow\downarrow)],\nonumber\\
\Xi^0_\uparrow&=&-\frac{1}{3\sqrt2} [uss\otimes(\uparrow\downarrow\uparrow-2\downarrow\uparrow\uparrow+\uparrow\uparrow\downarrow)+sus\otimes(\downarrow\uparrow\uparrow-2\uparrow\downarrow\uparrow+\uparrow\uparrow\downarrow) +ssu\otimes(\uparrow\downarrow\uparrow+ \downarrow\uparrow\uparrow-2\uparrow\uparrow\downarrow)],\nonumber\\
\Xi^-_\uparrow&=&\frac{1}{3\sqrt2} [sds\otimes(-2\uparrow\downarrow\uparrow+\downarrow\uparrow\uparrow+\uparrow\uparrow\downarrow)+dss\otimes(-2\downarrow\uparrow\uparrow+\uparrow\downarrow\uparrow+\uparrow\uparrow\downarrow) +ssd\otimes(\uparrow\downarrow\uparrow+ \downarrow\uparrow\uparrow-2\uparrow\uparrow\downarrow)],\nonumber\\
\Sigma^0_\uparrow&=&\frac16
[(dsu+usd)\otimes(2\uparrow\downarrow\uparrow-\downarrow\uparrow\uparrow-\uparrow\uparrow\downarrow)
+(sdu+sud)\otimes(2\downarrow\uparrow\uparrow-\uparrow\downarrow\uparrow-\uparrow\uparrow\downarrow) \nonumber\\
&&+(dus+uds)\otimes(-\uparrow\downarrow\uparrow- \downarrow\uparrow\uparrow+2\uparrow\uparrow\downarrow)],\nonumber\\
\Lambda_\uparrow&=&-\frac{1}{2\sqrt3}
[(dsu-usd)\otimes(\downarrow\uparrow\uparrow-\uparrow\uparrow\downarrow)
+(sdu-sud)\otimes(\uparrow\downarrow\uparrow-\uparrow\uparrow\downarrow)
+(uds-dus)\otimes(\uparrow\downarrow\uparrow- \downarrow\uparrow\uparrow)].
\end{eqnarray}
Note that we have added a minus sign for the wave functions of $\Sigma^+$, $\Xi^0$, and $\Lambda$. Without these additional signs, the relations $D=g_A^q$ and $F=\frac23g_A^q$ cannot be consistently obtained when one uses coupling vertices related to these baryons.

In our investigation, besides the calculation term by term, we also confirm the obtained relations with a computer program. To do that, one rewrites the above nucleon wave functions in two-flavor description as the form
\begin{eqnarray}
N^i_\alpha&=&\frac{1}{\sqrt{2}}[(U_S)^{i,xyz}(U_S)_{\alpha,\rho\tau\eta}+(U_A)^{i,xyz}(U_A)_{\alpha,\rho\tau\eta}]q^x_\rho q^y_\tau q^z_\eta
\equiv W^{i,xyz}_{\alpha,\rho\tau\eta}q^x_\rho q^y_\tau q^z_\eta,
\end{eqnarray}
where $i,x,y,z$ are flavor indices (1 represents $p$ or $u$ and 2 represents $n$ or $d$), $\alpha,\rho,\tau,\eta$ are spin indices (1 represents $\uparrow$ and 2 represents $\downarrow$), and the $U$ coefficients may be obtained with
\begin{eqnarray}
(U_S)_{i,xyz}=\frac{1}{\sqrt6}(\delta_{ix}\epsilon_{yz}+\delta_{iy}\epsilon_{xz}),\qquad
(U_A)_{i,xyz}=\frac{1}{\sqrt2}\epsilon_{xy}\delta_{iz}.
\end{eqnarray}
The $W$ coefficients can also be read out from Eq. \eqref{wf-octet} directly. It is easy to get the following properties
\begin{eqnarray}
W^{i,xyz}_{\alpha,\rho\tau\eta}&=&W^{i,xzy}_{\alpha,\rho\eta\tau}=W^{i,yxz}_{\alpha,\tau\rho\eta}=W^{i,yzx}_{\alpha,\tau\eta\rho}=W^{i,zyx}_{\alpha,\eta\tau\rho}=W^{i,zxy}_{\alpha,\eta\rho\tau},\\
\sum_{yz\tau\eta}W^{j,x'yz}_{\beta,\rho'\tau\eta}W^{i,xyz}_{\alpha,\rho\tau\eta}&=&
\frac59\delta^{ix}\delta^{jx'}\delta_{\alpha\rho}\delta_{\beta\rho'}+\frac29\delta^{ij}\delta^{xx'}
\delta_{\alpha\beta}\delta_{\rho\rho'}-\frac19(\delta^{ij}\delta^{xx'}\delta_{\alpha\rho}\delta_{\beta\rho'}
+\delta^{ix}\delta^{jx'}\delta_{\alpha\beta}\delta_{\rho\rho'}).
\end{eqnarray}
Similarly, one may rewrite the octet wave functions in the three-flavor description as
\begin{eqnarray}
B^{ij}_\alpha&=&\frac{1}{\sqrt{2}}[(T_S)^{ij,xyz}(U_S)_{\alpha,\rho\tau\eta}+(T_A)^{ij,xyz}(U_A)_{\alpha,\rho\tau\eta}] q^x_\rho q^y_\tau q^z_\eta
\equiv X^{ij,xyz}_{\alpha,\rho\tau\eta}q^x_\rho q^y_\tau q^z_\eta,
\end{eqnarray}
where $i,j,x,y,z=1,2,3$ ($\alpha,\rho,\tau,\eta=1,2$) are flavor (spin) indices and
\begin{eqnarray}
T_S^{ij,xyz}=\frac{1}{\sqrt6}(\delta^{ix}\epsilon^{yzj}+\delta^{iy}\epsilon^{xzj}),\qquad
T_A^{ij,xyz}=\frac{1}{\sqrt2}(\delta^{iz}\epsilon^{xyj}-\frac{1}{3}\delta^{ij}\epsilon^{xyz}).
\end{eqnarray}
The modified phases for $\Sigma^+$, $\Xi^0$, and $\Lambda$ have been counted in $T_S$ and $T_A$. The $X$ coefficients may also be read out from Eq. \eqref{wf-octet} directly. One has
\begin{eqnarray}
&&X^{ij,xyz}_{\alpha,\rho\tau\eta}=X^{ij,xzy}_{\alpha,\rho\eta\tau}=X^{ij,yxz}_{\alpha,\tau\rho\eta}=X^{ij,yzx}_{\alpha,\tau\eta\rho}=X^{ij,zyx}_{\alpha,\eta\tau\rho}=X^{ij,zxy}_{\alpha,\eta\rho\tau};\\
&&\sum_{yz\tau\eta}X^{ab,x'yz}_{\beta,\rho'\tau\eta}X^{ij,xyz}_{\alpha,\rho\tau\eta}\nonumber\\
&=&\frac59\delta^{ix}\delta^{jb}\delta^{ax'}\delta_{\alpha\rho}\delta_{\beta\rho'}
-\frac29(\delta^{ij}\delta^{ax'}\delta^{bx}+\delta^{ix}\delta^{jx'}\delta^{ab})\delta_{\alpha\rho}\delta_{\beta\rho'}
+\frac29(\delta^{ia}\delta^{jb}\delta^{xx'}-\delta^{ia}\delta^{jx'}\delta^{bx})\delta_{\alpha\beta}\delta_{\rho\rho'}
\nonumber\\
&&+\frac19(\delta^{ij}\delta^{ab}\delta^{xx'}+\delta^{ia}\delta^{jx'}\delta^{bx}-\delta^{ia}\delta^{jb}\delta^{xx'})\delta_{\alpha\rho}\delta_{\beta\rho'}\nonumber\\
&&+\frac19(\delta^{ix}\delta^{jx'}\delta^{ab}-\delta^{ix}\delta^{jb}\delta^{ax'}+\delta^{ij}\delta^{ax'}\delta^{bx}-\delta^{ij}\delta^{ab}\delta^{xx'})\delta_{\alpha\beta}\delta_{\rho\rho'}.
\end{eqnarray}

\section{LEC relations in the $SU(2)$ case}\label{sec4}

Since there is no difference between the flavor-spin structures of Lagrangians in $SU(2)_{\chi PT}$ and $SU(2)_{\chi QM}$, the LEC relations should be simple. At the leading order, the nonrelativistic forms of the coupling terms in $\chi$PT and $\chi$QM are
\begin{eqnarray}
{\cal L}_N&=&-\frac12g_A\psi_H^\dag\bm{\sigma}\cdot\bm{u}\psi_H=-\frac12g_A\psi_{H,\beta}^{j\dag}\bm{\sigma}_{\beta\alpha}\cdot\bm{u}^{ji}\psi_{H,\alpha}^i,\nonumber\\
{\cal L}_q&=&-\frac12g_A^q\psi_{q}^\dag\bm{\sigma}\cdot\bm{u}\psi_q=-\frac12g_A^q\psi_{q,\rho'}^{x'\dag}\bm{\sigma}_{\rho'\rho}\cdot\bm{u}^{x'x}\psi_{q,\rho}^x.
\end{eqnarray}
To be specific, we consider the $p_\uparrow$-$p_\uparrow$-$\pi^0$ coupling case. Then the matrix element at the hadron level is
\begin{eqnarray}
{\cal M}\sim g_A q_3\langle p_\uparrow|\sigma_3\tau_3|p_\uparrow\rangle=g_A q_3,
\end{eqnarray}
where $q_3$ is the $z$-component pion momentum and $\sigma_3/\tau_3$ is the third Pauli matrix. The calculation for the same vertex at the quark level is
\begin{eqnarray}
{\cal M}\sim g_A^q q_3\langle p_\uparrow|\sum_{i=1}^3\sigma^{(i)}_3\tau^{(i)}_3|p_\uparrow\rangle=\frac53g_A^q q_3,
\end{eqnarray}
where $i$ labels quarks and the wave function of $p_\uparrow$ in Eq. \eqref{wf-octet} is used. Therefore, we have $g_A=\frac53g_A^q$.

At the second chiral order, there are five groups of structures we need to consider. Extending the above procedure gives six coupling constant relations. Note that there are two structures in the fifth group. The description at the quark level and that at the hadron level for the coupling types may be not one to one. When one adopts $\chi_+$ instead of $\tilde{\chi}_+$ in the sixth term, the contributions of the two structures should be considered together and the resulting relations are $\alpha_6=\beta_6$ and $\alpha_7=\beta_6+3\beta_7$. When the traceless field $\tilde{\chi}_+$ is adopted, the resulting relations are simply $\alpha_6=\beta_6$ and $\alpha_7=3\beta_7$.

At the third chiral order, we have sixteen groups of LECs. Similar to the considerations at ${\cal O}(p^2)$, one gets 21 relations. All the obtained relations between $SU(2)_{\chi PT}$ and $SU(2)_{\chi QM}$ up to ${\cal O}(p^3)$ are summarized in Table \ref{SU2relations}. There are four types of LEC relations corresponding to respective nonrelativistic interacting structures
\begin{eqnarray}\label{correspondence-su2}
1\to \alpha_i=3\beta_i, \qquad \sigma\to\alpha_i=\beta_i,\qquad \tau\to\alpha_i=\beta_i,
\qquad \tau\otimes\sigma\to \alpha_i=\frac53\beta_i.
\end{eqnarray}
These results are consistent with the quark model calculation by using the Wigner-Eckart theorem \cite{Drechsel:1983hny}. Although high-order chiral Lagrangians involve more pion fields, no more types of LEC relations in the adopted approximation exist.

\begin{table}[htbp]
\caption{Coupling constant relations between $SU(2)_{\chi PT}$ and $SU(2)_{\chi QM}$.}\label{SU2relations}
\begin{tabular}{cccc}\hline\hline
&Group& $SU(2)_{\chi PT}\Leftrightarrow SU(2)_{\chi QM}$ \\\hline
${\cal O}(p^1)$ &1& $g_A=\frac53g_A^q$. \\\hline
${\cal O}(p^2)$ &1& $\alpha_1=3\beta_1$; \\
  &2&   $\alpha_2=\frac53\beta_2$;\\
  &3&   $\alpha_3=3\beta_3$;  \\
  &4&   $\alpha_4=\frac53\beta_4$; \\
  &$5$&   $\alpha_6=\beta_6$, $\alpha_7=3\beta_7$. \\\hline
${\cal O}(p^3)$ &1& $\alpha_1=\frac53\beta_1$, $\alpha_2=\frac53\beta_2$; \\
  &2&   $\alpha_3=3\beta_3$;\\
  &3&   $\alpha_4=\frac53\beta_4$, $\alpha_5=\frac53\beta_5$;\\
  &4&   $\alpha_6=\beta_6$;\\
  &5&   $\alpha_7=\beta_7$;\\
  &6&   $\alpha_8=\beta_8$;\\
  &7&   $\alpha_9=\frac53\beta_9$;\\
  &8&   $\alpha_{11}=3\beta_{11}$;\\
  &9&   $\alpha_{12}=\beta_{12}$;\\
  &10&   $\alpha_{14}=\frac53\beta_{14}$;\\
  &11&   $\alpha_{15}=\beta_{15}$;\\
  &12&   $\alpha_{16}=\beta_{16}$, $\alpha_{17}=\beta_{17}$;\\
  &13&   $\alpha_{18}=\frac53\beta_{18}$;\\
  &$14$&   $\alpha_{19}=\beta_{19}$, $\alpha_{20}=\frac53\beta_{20}$;\\
  &$15$&   $\alpha_{21}=\frac53\beta_{21}$, $\alpha_{22}=\beta_{22}$;\\
  &16&   $\alpha_{23}=\beta_{23}$.
\\\hline\hline
\end{tabular}
\end{table}

To confirm the obtained relations with a computer program, it is convenient to define the function (at leading order)
\begin{eqnarray}\label{G-fun}
G(i,\alpha\to j,\beta; \sigma,\tau)\equiv g_A (\sigma)_{\beta\alpha} (\tau)^{ji}
-3g^q_A\sum_{yz,\tau\eta} W^{j,x'yz}_{\beta,\rho'\tau\eta}(\sigma)_{\rho'\rho}(\tau)^{x'x}W^{i,xyz}_{\alpha,\rho\tau\eta}.
\end{eqnarray}
In studying a vertex at the quark level, $\sigma/\tau$ acts on the specific quark (e.q. , the third quark). From vanishing $G(i,\alpha\to j,\beta; \sigma,\tau)$, one gets the relation between $g_A$ and $g_A^q$. In the above example, one recovers the relation $g_A=\frac53g_A^q$ by setting $i=j=\alpha=\beta\to 1$, $\sigma\to\sigma_3$, and $\tau\to\tau_3$. Other choices of $i,j,\alpha,\beta,\sigma,\tau$ do not change the result. At high orders, one may define other $G$ functions in a similar way. Note that the unit matrix other than the Pauli matrix may be used in the definition, depending on the spin-flavor structures of the coupling terms. The relations in Table \ref{SU2relations} and the correspondences in Eq. \eqref{correspondence-su2} are easy to confirm by considering various couplings.

\section{LEC relations in the $SU(3)$ case}\label{sec5}

In this case, since the baryons are in the adjoint representation but the quarks are in the fundamental representation, there is no one-to-one correspondence between the quark-level description and the hadron-level description for a given coupling type. One should consider all the coupling terms in the same group together. We use the leading order Lagrangians as an example to illustrate the procedure. The nonrelativistic forms of the coupling terms read
\begin{eqnarray}
{\cal L}_B&\sim&(D+F)\langle B^\dag_H\bm{\sigma}\cdot \bm{u} B_H\rangle+(D-F)\langle B^\dag_H\bm{\sigma}\cdot B_H  \bm{u} \rangle,\notag\\
{\cal L}_q&\sim& g_A^q \Psi^\dag\bm{\sigma}\cdot\bm{u}\Psi.
\end{eqnarray}
The $p_\uparrow$-$p_\uparrow$-$\pi^0$ coupling and the $\Sigma^+_\uparrow$-$\Sigma^0_\uparrow$-$\pi^0$ coupling are determined by $D+F$ and $F$ at the hadron level, respectively, while they are both determined by $g_A^q$ at the quark level. The equivalence between hadron-level calculation and quark-level calculation gives $D+F=\frac53g_A^q$ and $F=\frac23g_A^q$, respectively. Therefore, we have $D=g_A^q$ and $F=\frac23g_A^q$, which is consistent with Ref. \cite{Jenkins:1991es}. Other coupling considerations do not change this result if the consistent wave functions in Eq. \eqref{wf-octet} are adopted.

Noticing the calculation difference between $SU(3)$ and $SU(2)$, we get relations between $c_i$ and $d_i$ at the second and third orders, which are collected in Table \ref{SU3relations}. From the results, relations between $d$'s can also be established. Some $d$'s are set to zero so that the LEC relations from different coupling vertices are consistent.

Since the situation in the $SU(3)$ case is more complicated than the $SU(2)$ case, the correspondences similar to Eq. \eqref{correspondence-su2} seem to be lost. In fact, from the fact that $SU(2)$ is a subgroup of $SU(3)$ and there are always two ways for the octet-octet-octet coupling ($\langle\bar{B}\lambda B\rangle$ and $\langle\bar{B}B\lambda \rangle$), one understands that some correspondences should exist. It is found that the LEC relations can be obtained with the following structure correspondences:
\begin{eqnarray}\label{correspondence-su3}
\langle\bar{B}B\rangle\to \bar{\Psi}\Psi &\Rightarrow& \text{(combination of several $d$'s)}=3\text{(combination of several $c$'s)},\nonumber\\
\langle\bar{B}\lambda B\rangle\to \bar{\Psi}\lambda\Psi &\Rightarrow& \text{(combination of several $d$'s)}=\text{(combination of several $c$'s)},\nonumber\\
\langle\bar{B}B\lambda\rangle\to \bar{\Psi}\lambda\Psi &\Rightarrow& \text{(combination of several $d$'s)}=-\text{(combination of several $c$'s)},\nonumber\\
\langle\bar{B}\lambda\sigma B\rangle\to\bar{\Psi}\lambda\sigma\Psi&\Rightarrow& \text{(combination of several $d$'s)}=\frac53\text{(combination of several $c$'s)},\nonumber\\
\langle\bar{B}\sigma B\lambda\rangle\to\bar{\Psi}\lambda\sigma\Psi&\Rightarrow& \text{(combination of several $d$'s)}=\frac13\text{(combination of several $c$'s)},\nonumber\\
\langle\bar{B}\sigma B\rangle\to\bar{\Psi}\sigma\Psi&\Rightarrow& \text{(combination of several $d$'s)}=\text{(combination of several $c$'s)}.
\end{eqnarray}
The first, second, fourth, and sixth correspondences are easy to understand. To get the third and fifth correspondences, one may study the couplings $\bar{B}_{13}(\lambda^3)_{11}B_{31}$ and $\bar{B}_{13}(\lambda^3)_{11}\sigma B_{31}$ or other couplings irrelevant with $\langle\bar{B}\lambda B\rangle$ and $\langle\bar{B}\lambda\sigma B\rangle$. The LECs for the structures of $\langle\bar{B}\lambda B\lambda\rangle$, $\langle\bar{B}\lambda\rangle\langle\lambda B\rangle$, $\langle\bar{B}\lambda\sigma B\lambda\rangle$, and $\langle\bar{B}\lambda\rangle\langle\sigma\lambda B\rangle$ can be set to zero in the adopted approximation. Before employing these correspondences, one needs to reorganize the Lagrangian terms in the same group with the help of the Cayley-Hamilton relations. For example, the complicated group-1 terms at ${\cal O}(p^3)$ should be reorganized to be
\begin{eqnarray}
&&(d_1-d_2+d_{11})\langle\bar{B}(\widetilde{u^\mu u_\mu u^\nu}+\widetilde{u^\nu u^\mu u_\mu})\gamma_\nu\gamma^5B\rangle
+(d_8-d_7+d_{11})\langle\bar{B}\gamma_\nu\gamma^5B(\widetilde{u^\mu u_\mu u^\nu}+\widetilde{u^\nu u^\mu u_\mu})\rangle\nonumber\\
&&+d_4\langle\bar{B}(\widetilde{u^\mu u^\nu}+\widetilde{u^\nu u^\mu})\gamma_\nu\gamma^5Bu_\mu\rangle
+d_6\langle\bar{B}u_\mu\gamma_\nu\gamma^5B(\widetilde{u^\mu u^\nu}+\widetilde{u^\nu u^\mu})\rangle\nonumber\\
&&+(d_3+d_{11})\langle\bar{B}\widetilde{u^\mu u_\mu}\gamma_\nu\gamma^5Bu^\nu\rangle +(d_5+d_{11})\langle\bar{B}u^\nu\gamma_\nu\gamma^5B\widetilde{u^\mu u_\mu}\rangle\nonumber\\
&&+(d_2+\frac23d_6)\langle\bar{B}u_\mu\gamma_\nu\gamma^5 B\rangle\langle u^\mu u^\nu\rangle
+(d_7+\frac23d_4)\langle\bar{B}\gamma_\nu\gamma^5 Bu_\mu\rangle\langle u^\mu u^\nu\rangle\nonumber\\
&&+(\frac12d_2+\frac13d_5-\frac23d_{11})\langle\bar{B}u^\nu\gamma_\nu\gamma^5B\rangle\langle u^\mu u_\mu\rangle
+(\frac13d_3+\frac12d_7+d_9-\frac23d_{11})\langle\bar{B}\gamma_\nu\gamma^5Bu^\nu\rangle\langle u^\mu u_\mu\rangle\nonumber\\
&&+(\frac23d_1+\frac13d_2+\frac13d_7+\frac23d_8+d_{10}+\frac13d_{11})\langle\bar{B}\gamma_\nu\gamma^5B\rangle\langle u^\mu u_\mu u^\nu\rangle,
\end{eqnarray}
where $\widetilde{X}$ indicates the traceless part of $X$. Then a set of LEC equations according to the above correspondences and thus the relations in Table \ref{SU3relations} can be obtained.

\begin{table}[htbp]
\caption{Coupling constant relations between $SU(3)_{\chi PT}$ and $SU(3)_{\chi QM}$.}\label{SU3relations}
\begin{tabular}{cccc}\hline\hline
&Group&$SU(3)_{\chi PT}\Leftrightarrow SU(3)_{\chi QM}$\\\hline
${\cal O}(p^1)$ &1& $D=g_A^q$, $F=\frac23g_A^q$. \\\hline
${\cal O}(p^2)$ &1& $d_1=-d_3=c_2$, $d_2=0$, $d_4=3c_1+c_2$; \\
  &2& $d_5=5d_6=\frac53c_3$, $d_7=0$;\\
  &3& $d_8=-d_{10}=c_5,d_9=0,d_{11}=3c_4+c_5$;\\
  &4& $d_{12}=5d_{13}=\frac53c_6$;\\
  &$5$& $d_{14}=-d_{15}=c_7$, $d_{16}=3c_8$.\\\hline
${\cal O}(p^3)$ &1& $d_1=\frac{5 }{6}(2c_1+c_2+2c_4)$, $d_2=5d_7=\frac53 c_2$, $d_3=d_5=-\frac54d_9=-d_{11}=\frac56(2c_1-c_2)$, \\
  &&  $d_4=d_6=0$, $d_8=\frac16(10c_1-3c_2+2c_4)$, $d_{10}=\frac16(-10c_1-7c_2+6c_3-4c_4)$;\\
  &2&  $d_{12}=-d_{15}=c_6$, $d_{13}=d_{14}=0$, $d_{16}=3c_5+c_6$; \\
  &3&  $d_{17}=\frac56(c_7+2c_8+2c_{10})$, $d_{18}=5d_{24}=\frac53 c_7$, $d_{19}=d_{22}=-\frac54d_{26}=-d_{27}=\frac56(-c_7+2c_8)$, \\
  &&  $d_{20}=d_{21}=0$, $d_{23}=\frac16(-3c_7+10c_8+2c_{10})$, $d_{25}=\frac16(-7c_7-10c_8+6c_9-4c_{10})$;\\
  &4&  $d_{28}=-d_{29}= c_{11}$, $d_{30}=0$;\\
  &5&  $d_{31}=-d_{32}= c_{12}$, $d_{33}=0$;\\
  &6&  $d_{34}=5d_{36}=\frac53 c_{14}$, $d_{35}=0$, $d_{37}=c_{13}-\frac23 c_{14}$;\\
  &7&  $d_{38}=5d_{39}=\frac53 c_{15}$, $d_{40}=0$;\\
  &8&  $d_{41}=-d_{44}= c_{17}$, $d_{42}=d_{43}=0$, $d_{45}=3c_{16}+2 c_{17}$;\\
  &9&  $d_{46}=-d_{47}= c_{18}$;\\
  &10&  $d_{48}=5d_{49}=\frac53 c_{19}$, $d_{50}=0$;\\
  &11&  $d_{51}=-d_{52}= c_{20}$, $d_{53}=0$;\\
  &12&  $d_{54}=5d_{60}=\frac53 c_{23}$, $d_{55}=5d_{61}=\frac53c_{24}$, $d_{56}=d_{57}=d_{58}=d_{59}=0$,\\
  &&  $d_{62}=c_{21}-\frac23c_{23}$, $d_{63}=c_{22}-\frac23 c_{24}$;\\
  &13&  $d_{64}=5d_{65}=\frac53 c_{25}$;\\
  &$14$&  $d_{66}=5d_{69}=\frac53 c_{28}$, $d_{67}=d_{68}=0$, $d_{70}=c_{26}-\frac23c_{28}$, $d_{71}=5d_{72}=\frac53c_{27}$;\\
  &$15$&  $d_{73}=5d_{74}=\frac53 c_{29}$, $d_{75}=c_{30}$;\\
  &16&  $d_{76}=-d_{77}=c_{31}$, $d_{78}=0$.
\\\hline\hline
\end{tabular}
\end{table}

To confirm the calculation with computer, one may define various $G$ functions similar to the $SU(2)$ case. At leading order, the definition is
\begin{eqnarray}
G(i,j,\alpha\to a,b,\beta; \sigma, \lambda)&\equiv&(\sigma)_{\beta\alpha}\left[(D+F)(\lambda)^{ai}\delta^{jb}+(D-F)\delta^{ai}(\lambda)^{jb}\right]\nonumber\\
&&-3g_A^q X_{\beta,\rho'\tau\eta}^{ab,x'yz}(\sigma)_{\rho'\rho}(\lambda)^{x'x}X_{\alpha,\rho\tau\eta}^{ij,xyz}.
\end{eqnarray}
From the vanishing $G$, the above results $D=g_A^q$ and $F=\frac23g_A^q$ are recovered by considering various couplings. However, one should note that $B_{11}$, $B_{22}$, and $B_{33}$ contain both $\Sigma^0$ and $\Lambda$. The combination of several $G$ functions is needed when the relevant couplings are involved. For example, if one extracts the relations from the $\Sigma^+$-$\pi^-$-$\Lambda$ coupling, $\Sigma^0$-$\pi^0$-$\Lambda$ coupling, and $\Lambda$-$\eta$-$\Lambda$ coupling, the equations we need to study should be
\begin{eqnarray}
\sum_x(\lambda_8)_{xx}G(1,2,\alpha\to x,x,\beta; \sigma, \lambda)&=&0,\nonumber\\
\sum_{x,y}(\lambda_3)_{xx}(\lambda_8)_{yy}G(x,x,\alpha\to y,y,\beta; \sigma, \lambda)&=&0,\nonumber\\
\sum_{x,y}(\lambda_8)_{xx}(\lambda_8)_{yy}G(x,x,\alpha\to y,y,\beta; \sigma, \lambda)&=&0,
\end{eqnarray}
respectively. By defining different $G$ functions at high orders and considering various coupling vertices, all the results in Table \ref{SU3relations} can be recovered.

\section{LEC relations in $\chi$QM and $\chi$PT}\label{sec6}

To find LEC relations between $SU(3)_{\chi PT}$ and $SU(2)_{\chi PT}$, we also need to know relations between $\beta_i$ and $c_i$. Now we consider this issue.

At the leading order, we adopted $g_A^q$ to denote the coupling constants in both $SU(3)_{\chi QM}$ and $SU(2)_{\chi QM}$ since there is only one coupling term. At the second and third orders, the situation is different. One picks up the $SU(2)$ sector in the Lagrangian of $SU(3)_{\chi QM}$ by replacing the field $\phi=\pi^i\lambda^i$ with $\phi=\pi^i\tau^i$ and using the Cayley-Hamiltonian relations for $2\times2$ matrices $X$, $Y$, and $Z$
\begin{eqnarray}\label{CHrelations}
YZ+ZY&=&Y\langle Z\rangle+ Z\langle Y\rangle +\langle YZ\rangle-\langle Y\rangle\langle Z\rangle,\nonumber\\
XYZ+YZX+ZXY&=&
\frac12\Big[\langle XY\rangle Z+\langle YZ\rangle X+\langle ZX\rangle Y
+XY\langle Z\rangle +YZ\langle X\rangle+ZX\langle Y\rangle\nonumber\\
&&
-\langle XY\rangle\langle Z\rangle-\langle YZ\rangle\langle X\rangle
-\langle ZX\rangle\langle Y\rangle
+3\langle XYZ\rangle\Big].
\end{eqnarray}
The matching between this sector and the Lagrangian of $SU(2)_{\chi QM}$ gives the coupling constant relations between $\beta$'s and $c$'s. We list all the obtained relations in Table \ref{chiQMrelations}. Not all the coupling constants in $SU(3)_{\chi QM}$ can be constrained by those in $SU(2)_{\chi QM}$. The ${\cal O}(p^3)$ terms $c_3\bar{\Psi}\la u^\mu u_\mu u^\nu\ra \gamma_\nu\gamma_5\Psi$ and $c_9\bar{\Psi}\la u^\mu u^\nu u^\lambda\ra\gamma_\mu\gamma_5  D_{\nu\lambda}\Psi$ always involve the $s$ quark contribution. We use $nc.$ to denote this case in the table.

\begin{table}[htbp]
\caption{Coupling constant relations between $SU(3)_{\chi QM}$ and $SU(2)_{\chi QM}$. The leading order coupling constants are both $g_A^q$. $nc.$ means that there is no constraint from the coupling constants in $SU(2)_{\chi QM}$.}\label{chiQMrelations}
\begin{tabular}{cccc}\hline\hline
&Group& $SU(3)_{\chi QM}\Leftrightarrow SU(2)_{\chi QM}$\\\hline
${\cal O}(p^1)$ &1& $g_A^q=g_A^q$. \\\hline
${\cal O}(p^2)$ &1& $\beta_1=c_1+\frac12c_2$; \\
  &2&   $\beta_2=c_3$;\\
  &3&   $\beta_3=c_4+\frac12c_5$;\\
  &4&   $\beta_4=c_6$;\\
  &$5$&   $\beta_6=c_7$, $\beta_7=\frac16 c_7+c_8$. \\\hline
${\cal O}(p^3)$ &1& $\beta_1=c_1+c_4$, $\beta_2=c_2$, $c_3$ ($nc.$); \\
  &2&   $\beta_3=c_5+\frac12c_6$;\\
  &3&   $\beta_4=c_7$, $\beta_5=c_8+c_{10}$, $c_9$ ($nc.$);\\
  &4&   $\beta_6=c_{11}$; \\
  &5&   $\beta_7=c_{12}$; \\
  &6&   $\beta_8=c_{13}+c_{14}$;\\
  &7&   $\beta_9=c_{15}$;  \\
  &8&   $\beta_{11}=c_{16}+c_{17}$;\\
  &9&   $\beta_{12}=c_{18}$;  \\
  &10&   $\beta_{14}=c_{19}$;  \\
  &11&   $\beta_{15}=c_{20}$;  \\
  &12&   $\beta_{16}=c_{21}+c_{23}$, $\beta_{17}=c_{22}+c_{24}$;  \\
  &13&   $\beta_{18}=c_{25}$;  \\
  &$14$&   $\beta_{19}=c_{26}+c_{28}$, $\beta_{20}=c_{27}+\frac13 c_{28}$;  \\
  &$15$&   $\beta_{21}=c_{29}$, $\beta_{22}=\frac{1}{6}c_{29}+ c_{30}$;  \\
  &16&   $\beta_{23}=c_{31}$.
\\\hline\hline
\end{tabular}
\end{table}

Combining the relations in Table \ref{chiQMrelations} with those in Tables \ref{SU2relations} and \ref{SU3relations}, one gets the final LEC relations between $SU(3)_{\chi PT}$ and $SU(2)_{\chi PT}$. They are shown in the third column of Table \ref{chiPTrelations}. All the relations are simple. There are LECs on which we cannot get constraints from the $SU(2)_{\chi PT}$. The processes that can constrain them must involve strange quark contributions.

\begin{table}[htbp]
\caption{LEC relations between $SU(2)_{\chi PT}$ and $SU(3)_{\chi PT}$. $nc.$ means that there is no constraint from the LECs in $SU(2)_{\chi PT}$.}\label{chiPTrelations}
\begin{tabular}{ccccc}\hline\hline
&Group& With $\chi QM$ & Without $\chi QM$ \\\hline
${\cal O}(p^1)$ &1& $D=\frac35g_A$, $F=\frac25g_A$. & $D+F=g_A$.\\
\hline
${\cal O}(p^2)$ &1& $\alpha_1=\frac{1}{2}d_1+d_4= -\frac12d_3+d_4$, $d_2=0$; & $\alpha_1=\dfrac{1}{2}d_1+d_4$; \\
  &2& $\alpha_2=d_5=5d_6$, $d_7=0$;   &  $\alpha_2=d_5$;\\
  &3& $\alpha_3=\frac{1}{2}d_8+d_{11}=-\frac{1}{2}d_{10}+d_{11}$, $d_9=0$;  & $\alpha_3=\frac12d_8+d_{11}$;\\
  &4& $\alpha_4=d_{12}=5d_{13}$;  & $\alpha_4=d_{12}$;\\
  &$5$& $\alpha_6=d_{14}=-d_{15}$, $\alpha_7=\frac12 d_{14}+d_{16}=-\frac12 d_{15}+d_{16}$.  &  $\alpha_6=d_{14}$, $\alpha_7=\frac12 d_{14}+d_{16}-\frac13(d_{14}+d_{15})$. \\
\hline
${\cal O}(p^3)$ &1& $\alpha_1+\frac12\alpha_2=d_1=5d_8-4d_3$, $d_3=d_5=-\frac54d_9=-d_{11}$,   & $\alpha_1=d_1-\frac12d_2$, $\alpha_2=d_2$;\\
&&$\alpha_2=d_2=5d_7$,   $d_4=d_6=0$,  $d_{10}$ (nc.);\\
  &2&  $\alpha_3=\frac{1}{2}d_{12}+d_{16}=-\frac{1}{2}d_{15}+d_{16}$, $d_{13}=d_{14}=0$; & $\alpha_3=\frac12d_{12}-d_{16}$; \\
  &3&  $\alpha_4=d_{18}=5d_{24}$, $\alpha_5+\frac12\alpha_4=d_{17}=5d_{23}-4d_{19}$,  &
    $\alpha_4=d_{18}$, $\alpha_5=d_{17}-\frac12d_{18}$;\\
  &&  $d_{19}=d_{22}=-\frac54d_{26}=-d_{27}$, $d_{20}=d_{21}=0$, $d_{25}$ ($nc.$);\\
  &4&  $\alpha_6=d_{28}=-d_{29}$, $d_{30}=0$;    &   $\alpha_6=d_{28}$;\\
  &5&  $\alpha_7=d_{31}=-d_{32}$, $d_{33}=0$;    &   $\alpha_7=d_{31}$;\\
  &6&  $\alpha_8=d_{34}+d_{37}=5d_{36}+d_{37}$, $d_{35}=0$;   &   $\alpha_8=d_{34}+d_{37}$;\\
  &7&  $\alpha_9=d_{38}=5d_{39}$, $d_{40}=0$;    &   $\alpha_9=d_{38}$;\\
  &8&  $\alpha_{11}=d_{41}+d_{45}=-d_{44}+d_{45}$, $d_{42}=d_{43}=0$;    &   $\alpha_{11}=d_{41}+d_{45}$;\\
  &9&  $\alpha_{12}=d_{46}=-d_{47}$;             &   $\alpha_{12}=d_{46}$;\\
  &10&  $\alpha_{14}=d_{48}=5d_{49}$, $d_{50}=0$; &   $\alpha_{14}=d_{48}$;\\
  &11&  $\alpha_{15}=d_{51}=-d_{52}$, $d_{53}=0$; &   $\alpha_{15}=d_{51}$;\\
  &12&  $\alpha_{16}=d_{54}+d_{62}=5d_{60}+d_{62}$, $\alpha_{17}=d_{55}+d_{63}=5d_{61}+d_{63}$,  &
    $\alpha_{16}=d_{54}+d_{62}$, $\alpha_{17}=d_{55}+d_{63}$;\\
  &&  $d_{56}=d_{57}=d_{58}=d_{59}=0$;\\
  &13&  $\alpha_{18}=d_{64}=5d_{65}$;   &   $\alpha_{18}=d_{64}$;  \\
  &$14$&  $\alpha_{19}=d_{66}+d_{70}=5d_{69}+d_{70}$, $d_{67}=d_{68}=0$, &$\alpha_{19}=d_{66}+d_{70}$, $\alpha_{20}=\frac13d_{66}-\frac13d_{68}+d_{71}$; \\
    &&$\alpha_{20}=\frac13d_{66}+d_{71}=\frac53d_{69}+d_{71}=\frac13d_{66}+5d_{72}$; \\
  &$15$&  $\alpha_{21}=d_{73}=5d_{74}$, $\alpha_{22}=\frac{1}{10}d_{73}+ d_{75}=\frac{1}{2}d_{74}+ d_{75}$;  &  $\alpha_{21}=d_{73}$, $\alpha_{22}=\frac{1}{6}d_{73}-\frac13d_{74}+d_{75}$; \\
  &16&  $\alpha_{23}=d_{76}=-d_{77}$, $d_{78}=0$.            &  $\alpha_{23}=d_{76}$.
\\\hline\hline
\end{tabular}
\end{table}

We also consider the LEC relations by picking up the $SU(2)_{\chi PT}$ terms in the $SU(3)_{\chi PT}$ Lagrangian, i.e. without using the $\chi QM$. They are helpful crosschecks for the results we obtain. From Tables \ref{p1}, \ref{p2}, and \ref{p3}, there are 6 flavor structures in the $SU(3)_{\chi PT}$ Lagrangian, $\langle \bar{B}YB\rangle$, $\langle\bar{B}BY\rangle$, $\langle\bar{B}YBZ\rangle$, $\langle\bar{B}B\rangle\langle Y\rangle$, $\langle \bar{B}Y\rangle \langle ZB\rangle$, and $\langle \bar{B}BY\rangle\langle Z\rangle$. It is not necessary to consider the following terms
\begin{eqnarray}
\langle \bar{B}B Y\rangle&=&\bar{B}_{ij}B_{jk}Y_{ki}
\to
\left(\begin{array}{cc}\bar{B}_{31}&\bar{B}_{32}\end{array}\right)
\left(\begin{array}{c}B_{13}\\B_{23}\end{array}\right)Y_{33},\\
\langle \bar{B}B Y\rangle\langle Z\rangle &=&\bar{B}_{ij}B_{jk}Y_{ki}\langle Z\rangle
\to
\left(\begin{array}{cc}\bar{B}_{31}&\bar{B}_{32}\end{array}\right)
\left(\begin{array}{c}B_{13}\\B_{23}\end{array}\right)Y_{33}\langle Z\rangle,\\
\langle \bar{B}YB Z\rangle&=&\bar{B}_{ij}Y_{jk}B_{kl}Z_{li}
\to
\left(\begin{array}{cc}\bar{B}_{31}&\bar{B}_{32}\end{array}\right)
\left(\begin{array}{cc}Y_{11}&Y_{12}\\Y_{21}&Y_{22}\end{array}\right)
\left(\begin{array}{c}B_{13}\\B_{23}\end{array}\right)Z_{33},
\end{eqnarray}
and
\begin{eqnarray}
\langle \bar{B}Y\rangle \langle ZB\rangle&=&\bar{B}_{ij}Y_{ji}B_{kl}Z_{lk}
\to
\left(\begin{array}{cc}\bar{B}_{31}&\bar{B}_{32}\end{array}\right)
\left(\begin{array}{c}Y_{13}\\Y_{23}\end{array}\right)
\left(\begin{array}{cc}Z_{31}&Z_{32}\end{array}\right)
\left(\begin{array}{c}B_{13}\\B_{23}\end{array}\right),
\end{eqnarray}
because they always involve $s$-quark contributions. The structure of the remaining terms one needs to consider is just like $\langle \bar{B}YB \rangle$ or $\langle \bar{B}B\rangle\langle Y\rangle$. Note that one should replace the $\widetilde{\chi}_+$ to $\chi_+-\frac{1}{N_f}\langle\chi_+\rangle$ and then let $Y=\chi_+$ when using this feature (similarly for $\widetilde{\chi}_-$ and $\widetilde{\chi}^\mu_\pm$).  Replacing the $3\times3$ matrix $B$ with the $2\times1$ matrix $\psi$ and the $3\times3$ pion matrices with the $2\times2$ pion matrices, one gets the needed terms. According to the Cayley-Hamilton relations in Eq. \eqref{CHrelations}, one finds the LEC relations listed in the fourth column of Table \ref{chiPTrelations}.

Let us move on to the comparison of the results obtained with and without using $\chi$QM. Obviously, the results with using $\chi$QM contain all the relations obtained without using $\chi$QM and the quark model symmetry leads to more LEC relations. This is easy to understand by analyzing the ${\cal O}(p^1)$ relations $D=\frac35g_A$ and $F=\frac25g_A$. In the method without $\chi$QM, the extraction of $D+F=g_A$ does not involve hyperon interactions. In the method with $\chi$QM, however, the hyperon interactions are considered to give more constraints in the approximation that the $s$ quarks are just spectators.

In the large $N_c$ limit, one may also derive some relations between the LECs. Noticing the symbol difference, we find that the relations $F/D=2/3$ at ${\cal O}(p^1)$ and $d_{30}=d_{33}=d_{35}=0$, $d_{31}=-d_{32}$, $d_{34}=5d_{36}$, $d_{73}=5d_{74}$, $d_{67}=d_{68}=0$, and $d_{66}-5d_{69}=\frac32(d_{71}-5d_{72})$ at ${\cal O}(p^3)$ are consistent with the large $N_c$ analysis performed in Refs. \cite{Heo:2019cqo,Heo:2022huf}.

\section{Numerical analysis}\label{sec7}

The LEC relations obtained in Table \ref{chiPTrelations} are approximate results. They are certainly affected by symmetry breaking and corrections from low order Lagrangians. An example is the matching relations at ${\cal O}(p^2)$ presented in Ref. \cite{Mai:2009ce}. It is helpful for us to check our approximate relations by taking a look at some numerical values. Now we consider this issue order by order.

\subsection{${\cal{O}}(p^1)$ LECs}

The $SU(2)$ coupling constant $g_A$ is extracted from the neutron beta decay and we take
$g_A=1.2694\pm0.0028$ here \cite{Scherer:2012zzc}. The $SU(3)$ LECs $D$ and $F$ are determined to be 0.80 and 0.50, respectively, by fitting the semileptonic decays $B\rightarrow B^{'} +e^- +{\bar{\nu}}_e$ \cite{Borasoy:1998pe}. Recently, a lattice calculation gives $D=0.730^{(11)}_{(11)}$, $F=0.447^{(6)}_{(7)}$, and $F/D=0.612^{(14)}_{(12)}$ \cite{Bali:2022qja}. These results confirm the well-known relation $g_A=D+F$. The relations $D=\frac35g_A$, $F=\frac25g_A$, and $F/D=2/3$ are also roughly satisfied.

\subsection{${\cal{O}}(p^2)$ LECs}

The LEC names we adopt differ from those in the literature. It is necessary to set up the relations between our LECs and those in the literature when using their values. In the $SU(2)$ case, we take the values from Ref. \cite{Alarcon:2012kn} which are compatible with those in Refs. \cite{Bernard:1996gq,Fettes:1998ud}. In the $SU(3)$ case, we refer to \cite{Huang:2019not}. The investigations of pion-nucleon scatterings in these two papers are both conducted to the third chiral order. With the help of the Cayley-Hamilton relation for traceless $3\times 3$ matrices given in Ref. \cite{Muller:1996vy}
\begin{eqnarray}
\langle\bar{B}u^\mu\rangle\langle B u_\mu\rangle=\langle\bar{B}u^\mu Bu_\mu\rangle+\langle\bar{B}B u_\mu u^\mu \rangle +\langle \bar{B}u^\mu u_\mu B\rangle -\frac12\langle\bar{B}B\rangle\langle u^\mu u_\mu\rangle,
\end{eqnarray}
we get the correspondences
\begin{center}
\begin{tabular}{lcl}
SU(2)&&SU(3)\\
$c_3\to 2\alpha_1=-6.74(38)\,\rm{GeV}^{-1}$, &&  $C_1\to d_1+2d_4=-6.75\pm0.14\,\rm{GeV}^{-1}$;\\
$c_4\to 2\alpha_2=3.74(16)\,\rm{GeV}^{-1}$,  && $C_3\to d_5=1.57\pm0.06\,\rm{GeV}^{-1}$;\\
$c_2\to -4\alpha_3m^2=4.08(19)\,\rm{GeV}^{-1}$,&& $C_2\to -2(d_8+2d_{11})m^2=5.30\pm0.35\,\rm{GeV}^{-1}$;\\
$c_1\to\alpha_7=-1.26(14)\,\rm{GeV}^{-1}$,&&$C_0\to d_{14}+2d_{16}-\frac23(d_{14}+d_{15})=-1.79\pm0.30\,\rm{GeV}^{-1}$;\\
&&$b_D+b_F\to d_{14}=-0.42\pm0.0\,\rm{GeV}^{-1}$;\\
&&$b_D-b_F\to d_{15}=0.54\pm0.0\,\rm{GeV}^{-1}$.\\
\end{tabular}
\end{center}
at ${\cal O}(p^2)$, where $m$ is the baryon mass in the chiral quark limit. One sees that the relations $2\alpha_1=d_1+2d_4$, $\alpha_2=d_5$, $2\alpha_3=d_8+2d_{11}$, $2\alpha_7=d_{14}+2d_{16}$, and $d_{14}=-d_{15}$ are roughly satisfied.

In Ref. \cite{Lu:2018zof}, the authors also studied the pion-nucleon scatterings up to the third chiral order. The numerical values with our notations are 
$d_1+2d_4=-7.63(6)\,\rm{GeV}^{-1}$, 
$\frac12d_5=1.34(1)\,\rm{GeV}^{-1}$, 
$-m(\frac12d_8+d_{11})=1.42(2)\,\rm{GeV}^{-2}$, and 
$\frac12d_{14}+d_{16}-\frac13(d_{14}+d_{15})=-1.36(6)\,\rm{GeV}^{-1}$. The signs are all consistent but the numerical deviations for the relations $\alpha_2=d_5$ and $2\alpha_7=d_{14}+2d_{16}$ are slightly larger. The relativistic correction probably has larger effects on $d_5$ and $\frac12d_{14}+d_{16}$.

In the same paper, the kaon-nucleon scatterings are also explored. Since the ${\cal O}(p^3)$ LECs are not included, we just take a look at the relations to the second order. From the LEC combinations and the numerical values, one gets
\begin{center}
\begin{tabular}{l}
SU(3)\\
$\alpha_1-\frac12\beta_1-\frac32\gamma_1\to -\frac12d_3+d_4= -7.20(6) \,\rm{GeV}^{-1}$;\\
$\beta_3-\gamma_3 \to \frac12d_6=0.183(2)\,\rm{GeV}^{-1}$;\\
$2\alpha_4+\frac12\beta_4-\frac32\gamma_4\to-\frac12d_{15}+d_{16}-\frac13(d_{14}+d_{15})=-2.25(12)\,\rm{GeV}^{-1}$;\\
$-2\alpha_2+\beta_2+3\gamma_2\to m(-\frac12d_{10}+d_{11})=-2.70(4)\,\rm{GeV}^{-1}$;\\
$-\beta_4+\gamma_4\to d_{14}=-0.406(1)\,\rm{GeV}^{-1}$;\\
$-2\alpha_4+2\gamma_4\to d_{15}=2.18(12) \,\rm{GeV}^{-1}$.
\end{tabular}
\end{center}
Obviously, only $d_5=5d_6$ and the signs for the relations $d_1+2d_4=-d_3+2d_4$, $d_8+2d_{11}=-d_{10}+d_{11}$, and $d_{14}=-d_{15}$ are confirmed. However, it does not mean that the LEC relations we obtain are incorrect since the values of ${\cal O}(p^2)$ LECs may be changed after the ${\cal O}(p^3)$ LECs are determined with enough experimental data.

\subsection{${\cal{O}}(p^3)$ LECs}

At this order, the comparison between the LECs in Refs. \cite{Alarcon:2012kn,Huang:2019not} and ours gives
\begin{center}
\begin{tabular}{lcl}
SU(2)&&SU(3)\\
$d_1+d_2\to -2m\alpha_6=3.3(7)\,\rm{GeV}^{-2}$,&&$H_2\to -2md_{28}=4.68\pm0.23\,\rm{GeV}^{-2}$;\\
$d_5\to -m\alpha_{23}=0.50(35)\,\rm{GeV}^{-2}$,&& $H_1\to -4m(d_{28}+d_{76})=4.84\pm2.57\,\rm{GeV}^{-2}$;\\
$-2m(\alpha_6+\alpha_{23})=4.3(1.0)\,\rm{GeV}^{-2}$,&& $-4m d_{76}=-4.52\pm2.61\,\rm{GeV}^{-2}$;\\
$d_3\to 12m^3\alpha_7=-2.7(6)\,\rm{GeV}^{-2}$,&& $H_3\to 24m^3d_{31}=-6.71\pm2.12\,\rm{GeV}^{-2}$;\\
$d_{14}-d_{15}\to -4m\alpha_8=-6.1(1.2)\,\rm{GeV}^{-2}$,&&$H_4\to -4m(d_{34}+d_{37})=-6.69\pm0.57\,\rm{GeV}^{-2}$.
\end{tabular}
\end{center}
With these numbers, one finds that the relations $\alpha_6=d_{28}$, $\alpha_7=d_{31}$, $\alpha_8=d_{34}+d_{37}$, and $\alpha_6+\alpha_{23}=d_{28}+d_{76}$ are roughly satisfied, but the relation $\alpha_{23}=d_{76}$ cannot be confirmed. Note that the loop corrections from kaon and eta have been included in the study of pion-nucleon scattering in Ref. \cite{Huang:2019not}. Such corrections probably affect largely on the value of $d_{76}$.

From Ref. [71], one gets 
$-md_{28}= 3.25(5)\,\rm{GeV}^{-2}$, 
$-\frac32md_{31} = 0.61(2)\,\rm{GeV}^{-2}$, 
$\frac12(d_{34}+d_{37})=1.45(3)\,\rm{GeV}^{-2}$, 
$-md_{76}=-0.32(13)\,\rm{GeV}^{-2}$, and $-m(d_{28}+d_{76})=2.93(14)\,\rm{GeV}^{-2}$. Of the relations $\alpha_6=d_{28}$, $\alpha_7=d_{31}$, $\alpha_8=d_{34}+d_{37}$, and $\alpha_6+\alpha_{23}=d_{28}+d_{76}$, the signs are all consistent and $\alpha_6+\alpha_{23}=d_{28}+d_{76}$ is roughly satisfied. Again, the relation $\alpha_{23}=d_{76}$ is not confirmed. Compared to the heavy baryon formalism \cite{Huang:2019not}, it seems that the relativistic correction has larger effects on the relations.

\section{Discussions}\label{sec8}

At the quark level, there are no structures similar to the ${\cal O}(p^3)$ $c_3$ and $c_9$ terms in the $SU(2)_{\chi QM}$. The coupling constants $c_3$ and $c_9$ do not get constraints from $SU(2)_{\chi QM}$, but it does not mean $c_3=c_9=0$. These two terms involve $s$-quark contributions. A direct consequence is that the extraction of ${\cal O}(p^3)$ $d_{10}$ and $d_{25}$ must rely on $s$-quark related processes. In our approximation (Fig. \ref{chiQM}), some LECs in $SU(3)_{\chi PT}$ are found to be zero. This means that such terms are negligible in some calculations when the available experimental data are not enough to determine all the LECs.

\begin{figure}[htbp]
\centering
\includegraphics{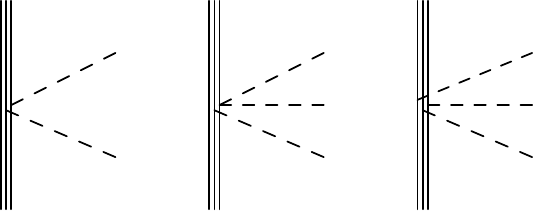}
\caption{Possible corrections to a baryon-baryon-meson coupling at the quark level.}\label{chiQM2}
\end{figure}

The obtained relations are just approximately correct. They are affected by flavor symmetry breaking and chiral corrections. We have assumed that the coupling of the pion fields with a baryon is described by the coupling of the pions with only one quark inside the baryon. If the couplings with different quarks are considered as shown in Fig. \ref{chiQM2}, the obtained LEC relations should also be improved. We leave the discussions in a future work. In the adopted chiral Lagrangians, the external sources $v_\mu$ and $a_\mu$ are traceless. If terms with $\la v_\mu \rangle\neq0,\langle a_\mu\rangle\neq0$ are constructed, one can find more relations and some relations would be revised accordingly.

The present study  involves at most ${\cal O}(p^3)$ chiral Lagrangians. At higher orders, the LEC relations between $\chi$PT and $\chi$QM are not difficult to get according to Eqs. \eqref{correspondence-su2} and \eqref{correspondence-su3}, once the required $\chi$QM Lagrangians are constructed. The LEC relations between $SU(2)_{\chi PT}$ and $SU(3)_{\chi PT}$ can also be obtained. In those cases, because the number of terms is increased, more $SU(3)$ LECs cannot be constrained by the $SU(2)$ LECs. The uncertainties in numerical results should be larger and the consideration of corrections to the relations would be more essential.

To summarize, we obtain some LEC relations (Table \ref{chiPTrelations}) between $SU(2)_{\chi PT}$ and $SU(3)_{\chi PT}$ at the orders ${\cal O}(p^1)$, ${\cal O}(p^2)$, and ${\cal O}(p^3)$ by employing the quark model symmetry in the approximation illustrated in Fig. \ref{chiQM}. The LEC relations between different $SU(3)_{\chi PT}$ terms at the same order are also found concurrently. The study in this work gives some vanishing LECs. The numerical analysis confirms our results to some extent. With such relations, the number of LECs to be determined from the experimental data can be reduced.

\section*{Acknowledgments}

Y.R.L thanks Professors Shi-Lin Zhu, Makoto Oka, Fei Huang (UCAS), and Li-Sheng Geng for helpful discussions. This project was supported by the National Natural Science Foundation of China under Grants No. 11775130,  No. 11775132, No. 11875179, and No. 11905112 and Guangxi Science Foundation under Grants No. 2022GXNSFAA035489 and 2018GXNSFAA281180.


\end{document}